\documentclass[twocolumn,showpacs,preprintnumbers,amsmath,amssymb]{revtex4}

\usepackage{graphicx}
\usepackage{dcolumn}
\usepackage{bm}
\usepackage{rotating}
\usepackage{color}

\newcommand{\target}{{*}}

\newcommand{\kslash}{k\kern-1ex /}
\newcommand{\pslash}{p\kern-1ex /}
\newcommand{\qslash}{q\kern-1ex /}
\newcommand{\lslash}{l\kern-1ex /}
\newcommand{\sslash}{s\kern-1ex /}
\newcommand{\Dslash}{D\kern-1.2ex /}

\newcommand{\tr}{{\rm tr}}

\newcommand{\beqa}{\begin{eqnarray}}
\newcommand{\eeqa}{\end{eqnarray}}

\newcommand{\bd}{\begin{description}}
\newcommand{\ed}{\end{description}}
\newcommand{\la}{\langle}
\newcommand{\ra}{\rangle}
\newcommand{\ben}{\begin{eqnarray}}
\newcommand{\een}{\end{eqnarray}}
\newcommand{\nn}{\nonumber}
\def\lsim{\raise0.3ex\hbox{$<$\kern-0.75em\raise-1.1ex\hbox{$\sim$}}}
\def\gsim{\raise0.3ex\hbox{$>$\kern-0.75em\raise-1.1ex\hbox{$\sim$}}}
\def\simgt{\rlap{\lower 3.5 pt\hbox{$\mathchar \sim$}}\raise 2.0pt \hbox {$>$}}
\def\simlt{\rlap{\lower 3.5 pt\hbox{$\mathchar \sim$}}\raise 2.0pt \hbox {$<$}}

\newcommand{\msbar}{{\overline {\rm MS}}}

\newcommand{\csw}{{c_{\rm SW}}}

\begin{document}

\preprint{UTCCS-P-56, UTHEP-597, HUPD-0908}

\title{Physical Point Simulation in 2+1 Flavor Lattice QCD}

\author{
 S.~Aoki${}^{a,b}$,
 K.-I.~Ishikawa${}^{d}$,
 N.~Ishizuka${}^{a,c}$,
 T.~Izubuchi${}^{b,e}$,
 D.~Kadoh${}^{c}$\footnote{Present address: Theoretical Physics Laboratory, 
RIKEN, Wako 2-1, Saitama 351-0198, Japan},
 K.~Kanaya${}^{a}$,
 Y.~Kuramashi${}^{a,c}$,
 Y.~Namekawa${}^{c}$,
 M.~Okawa${}^{d}$,
 Y.~Taniguchi${}^{a,c}$,
 A.~Ukawa${}^{a,c}$,
 N.~Ukita${}^{c}$,
 T.~Yamazaki${}^{c}$,
 T.~Yoshi\'e${}^{a,c}$\\
(PACS-CS Collaboration)
}
\affiliation{
 ${}^a$Graduate School of Pure and Applied Sciences, University of Tsukuba, Tsukuba, Ibaraki 305-8571, Japan\\
 ${}^b$Riken BNL Research Center, Brookhaven National Laboratory, Upton,
New York 11973, USA\\
 ${}^c$Center for Computational Sciences, University of Tsukuba, Tsukuba, Ibaraki 305-8577, Japan\\
 ${}^d$Graduate School of Science, Hiroshima University, Higashi-Hiroshima, Hiroshima 739-8526, Japan\\
 ${}^e$Institute for Theoretical Physics, Kanazawa University, Kanazawa,
       Ishikawa 920-1192, Japan
}



\date{\today}

\begin{abstract}
We present the results of the physical point simulation in 2+1 flavor
lattice QCD with the nonperturbatively $O(a)$-improved
Wilson quark action and the Iwasaki gauge action at $\beta=1.9$ on a
 $32^3\times 64$ lattice. The physical quark masses together with 
the lattice spacing is determined with 
$m_\pi$, $m_K$ and $m_\Omega$ as physical inputs. 
There are two key algorithmic ingredients
to make possible the direct simulation at the physical point: 
One is the mass-preconditioned domain-decomposed HMC algorithm to reduce
the computational cost. The other is the reweighting technique to
adjust the hopping parameters exactly to the physical point. 
The physics results include the hadron spectrum, the quark masses
and the pseudoscalar meson decay constants. The
renormalization factors are nonperturbatively evaluated 
with the Schr{\"o}dinger functional method.
The results are compared with the previous ones obtained 
by the chiral extrapolation method. 
\end{abstract}

\maketitle

\section{INTRODUCTION}
\label{sec:intro}

The physical point simulation is one of the essential ingredients
in the first principle calculation of lattice QCD. 
However, it is still a tough challenge because of the rapid growth of 
the computational cost with 
the up-down (ud) quark mass reduced toward its physical value.
At present simulation points are typically restricted to $m_\pi\simgt
250$ MeV. The most popular strategy to obtain the results at the
physical point is chiral extrapolation with the use 
of chiral perturbation theory (ChPT) as a guiding principle.
This strategy, however, has several problems:
(i) It is numerically difficult to precisely trace the logarithmic 
quark mass dependence of the physical quantities predicted by ChPT.
(ii) It may not be always possible to resort to ChPT as a good guiding
principle for chiral extrapolation.
(iii) The kinematics changes as the quark mass increases.  
A typical example is the $\rho\rightarrow \pi\pi$ decay which is not allowed
for the increased ud quark mass away from the physical value.
(iv) Our final destination is to incorporate the different up and down
quark masses. The isospin breaking effects are so tiny that
reliable evaluation would be difficult by the chiral extrapolation method.

In this article we present the results of the physical point simulation 
which has been pursued as the PACS-CS project 
based on the PACS-CS (parallel array computer system for 
computational sciences) computer with a peak speed of 14.3 Tflops
developed at University of Tsukuba\cite{ukawa1,ukawa2,boku}.
The simulation is carried out with the nonperturbatively $O(a)$-improved 
Wilson quark action\cite{csw}
and the Iwasaki gauge action\cite{iwasaki} on a (3 fm$)^3$ box at the lattice
spacing of $a=0.08995(40)$ fm.
There are two types of problems in the physical point simulation.
First, we need to reduce the computational cost which rapidly increases
as the ud quark mass decreases. This difficulty is overcome thanks to
the domain-decomposed HMC (DDHMC) algorithm\cite{luscher} 
with the mass-preconditioning\cite{massprec1,massprec2}. 
In Refs.~\cite{pacscs_nf3,pacscs_hbchpt} 
this algorithm was successfully applied to 
investigate the chiral behaviors of the pseudoscalar meson
sector and the hadron masses including both the mesons and the baryons,
where the pion mass covers from 156 MeV to 702 MeV.
The second problem is fine-tuning of the quark masses to the
physical point after we reach around the physical point.
This task is accomplished with the reweighting technique which allows
us to cover a small variation of simulation parameters 
in a single Monte Carlo run\cite{reweighting}.
We explain the details of the method and present the physics results
on the physical point without interpolation or extrapolation.

This paper is organized as follows.
In Sec.~\ref{sec:detail} we present the simulation details
including the parameters and the algorithm. 
Section~\ref{sec:reweighting} is devoted to describe the reweighting method.
We present the physics results on the physical point 
in Sec.~\ref{sec:hadron}.
Our conclusions are summarized in Sec.~\ref{sec:conclusion}.

\section{SIMULATION DETAILS}
\label{sec:detail}

\subsection{Actions}
\label{subsec:action}

We employ the Iwasaki gauge action\cite{iwasaki} and 
the nonperturbatively $O(a)$-improved Wilson quark action
as in the previous works\cite{pacscs_nf3,cppacs/jlqcd_nf3}.
The former is composed of a plaquette and a $1\times 2$ rectangle loop:
\ben
S_{\rm g}=\frac{1}{g^2}\left\{ c_0\sum_{\rm plaquette}\tr U_{pl}
+c_1\sum_{\rm rectangle}\tr U_{rtg} \right\}
\label{eq:action_g}
\een
with $c_1=-0.331$ and $c_0=1-8c_1=3.648$.
The latter is expressed as
\begin{widetext}
\ben
S_{\rm quark}&=&\sum_{q={\rm  u,d,s}}\left[
\sum_n {\bar q}_n q_n 
    -\kappa_q \csw \sum_n \sum_{\mu,\nu}\frac{i}{2}
              {\bar q}_n\sigma_{\mu\nu}F_{\mu\nu}(n)q_n
\right.\nn\\
&&\left.
-\kappa_q\sum_n
\sum_\mu \left\{{\bar q}_n(1-\gamma_\mu)U_{n,\mu} q_{n+{\hat \mu}}
               +{\bar q}_n(1+\gamma_\mu)U^{\dag}_{n-{\hat \mu},\mu} q_{n-{\hat \mu}}\right\}
 \right],
\label{eq:action_q}
\een
\end{widetext}
where we consider the case of a degenerate up and down quark mass
$\kappa_{\rm u}=\kappa_{\rm d}$.
The Euclidean gamma matrices are defined in terms of
the Minkowski matrices in the Bjorken-Drell convention:
$\gamma_j=-i\gamma_{BD}^j$ $(j=1,2,3)$, 
$\gamma_4=\gamma_{BD}^0$,
$\gamma_5=\gamma_{BD}^5$ and 
$\sigma_{\mu\nu}=\frac{1}{2}[\gamma_\mu,\gamma_\nu]$.
The field strength $F_{\mu\nu}$ in the clover term
is given by
\ben 
F_{\mu\nu}(n)&=&\frac{1}{4}\sum_{i=1}^{4}\frac{1}{2i}
\left(U_i(n)-U_i^\dagger(n)\right), \\
U_1(n)&=&U_{n,\mu}U_{n+{\hat \mu},\nu}
         U^\dagger_{n+{\hat \nu},\mu}U^\dagger_{n,\nu}, \\
U_2(n)&=&U_{n,\nu}U^\dagger_{n-{\hat \mu}+{\hat \nu},\mu}
         U^\dagger_{n-{\hat \mu},\nu}U_{n-{\hat \mu},\mu}, \\
U_3(n)&=&U^\dagger_{n-{\hat \mu},\mu}U^\dagger_{n-{\hat \mu}-{\hat \nu},\nu}
         U_{n-{\hat \mu}-{\hat \nu},\mu}U_{n-{\hat \nu},\nu}, \\
U_4(n)&=&U^\dagger_{n-{\hat \nu},\nu}U_{n-{\hat \nu},\mu}
         U_{n+{\hat \mu}-{\hat \nu},\nu}U^\dagger_{n,\mu}.
\een
The improvement coefficient $\csw$ for $O(a)$ improvement was 
determined nonperturbatively in Ref.~\cite{csw}.

\subsection{Simulation parameters}
\label{subsec:params}

Simulations are performed employing the same parameters
as in the previous work\cite{pacscs_nf3}:
a $32^3\times 64$ lattice at  $\beta=1.90$ with  $\csw=1.715$~\cite{csw}.
We choose 
$(\kappa_{\rm ud},\kappa_{\rm s})=(0.137\, 785\, 00,0.136\, 600\, 00)$
for a degenerate pair of up and down quarks and a strange quark.
This combination of the hopping parameters 
was supposed to be the physical point
based on the analysis of the previous results\cite{pacscs_nf3}.
The lattice spacing is determined as $a=0.08995(40)$ fm from  
the $m_\pi,m_K,m_\Omega$ results 
on the physical point after the reweighting procedure.
Table~\ref{tab:param} summarizes the run parameters.
After thermalization we calculate 
hadronic observables solving quark
propagators at every 20 trajectories (5 MD time units), 
while we measure the plaquette
expectation value at every trajectory.
The reweighting factors for the up-down and the strange quarks 
are evaluated at every 100 trajectories (25 MD time units). 
The choice of sparse measurements
is due to the demanding computational cost of the reweighting factors.  
The hadronic observables measured at every 20 and 100 trajectories
show consistency within error bars.
For the pion mass we find that the former 
has larger magnitude of error:
0.0719(37) and 0.0693(27). 
This could be due to wavy behavior of pion propagators on a couple of 
configurations caused by the statistical fluctuation.

\subsection{Algorithm}
\label{subsec:algorithm}

Our base algorithm for the degenerate up-down quarks is the
DDHMC algorithm\cite{luscher} which makes a geometric separation of 
the up-down quark determinant into the UV and the IR parts with
the domain-decomposition of the full lattice into small blocks.   
This UV/IR separation naturally introduces the multiple time integration
scheme\cite{sexton} in the molecular dynamics (MD) steps.
We employ the nested simple leapfrog with QPQ ordering for the multiple time 
step MD integrator.
According to the relative magnitude of the force terms coming from
the gauge part and the UV and the IR parts of the up-down quarks
we choose the associated step sizes such that
\ben
 \delta\tau_{\rm g} \|F_{\rm g}\| \approx \delta\tau_{\rm UV} \|F_{\rm UV}\| \approx \delta\tau_{\rm IR} \|F_{\rm IR}\|,
\een 
where $\delta\tau_{\rm g}=\tau/N_0 N_1 N_2,\ \  \delta\tau_{\rm UV}=\tau/N_1
N_2,\ \  \delta\tau_{\rm IR}=\tau/N_2$ with $\tau$ the trajectory
length and $(N_0,N_1,N_2)$ a set of integers to control the step sizes.

In the previous work we used the mass-preconditioned DDHMC
(MPDDHMC) algorithm for the run 
at $(\kappa_{\rm ud},\kappa_{\rm s})=(0.137\, 810\, 00,0.136\, 400\, 00)$ 
which gives the lightest up-down quark mass\cite{pacscs_nf3}.
A preconditioner controlled by an additional hopping parameter 
$\kappa_{\rm ud}^\prime=\rho\kappa_{\rm ud}$ 
is incorporated to tame the fluctuation of the IR force
$F_{\rm IR}$ in the original DDHMC algorithm by dividing
it into ${\tilde F}_{\rm IR}$ and $F^\prime_{\rm IR}$.
The former is derived from the preconditioned action and the latter from
the preconditioner. 
In this work we employ 
twofold-mass-preconditioned DDHMC (MP$^2$DDHMC) algorithm
which split $F_{\rm IR}$ into ${\tilde F}_{\rm IR}$, $F^\prime_{\rm IR}$ and 
$F^{\prime\prime}_{\rm IR}$.
This decomposition is controlled by two additional hopping parameters
$\kappa^\prime_{\rm ud}=\rho_1\kappa$ and 
$\kappa^{\prime\prime}_{\rm ud}=\rho_1\rho_2\kappa$.
${\tilde F}_{\rm IR}$ is derived from the action preconditioned with
$\kappa^\prime_{\rm ud}$. The ratio of two preconditioners with 
$\kappa^\prime_{\rm ud}$ and $\kappa^{\prime\prime}_{\rm ud}$ gives
$F^{\prime}_{\rm IR}$.  
$F^{\prime\prime}_{\rm IR}$ is from the heaviest preconditioners 
with $\kappa^{\prime\prime}_{\rm ud}$.
We find the following relative magnitude for the force terms:
\begin{widetext}
\ben
\|F_{\rm g}\|:\|F_{\rm UV}\|:\|F_{\rm IR}^{\prime\prime}\|:\|F_{\rm IR}^\prime\|:\|{\tilde F}_{\rm IR}\| \approx 16:4:1:1/7:1/60
\label{eq:force_mp2ddhmc}
\een
\end{widetext}
with $\rho_1=0.9995$ and $\rho_2=0.9900$.
We choose $(N_0,N_1,N_2,N_3,N_4)=(4,4,2,4,4)$ for the associated step
sizes: $\delta\tau_{\rm g}=\tau/N_0 N_1 N_2 N_3 N_4,$ 
$\delta\tau_{\rm UV}=\tau/N_1 N_2 N_3 N_4,$ 
$\delta\tau^{\prime\prime}_{\rm IR}=\tau/N_2 N_3 N_4,$ 
$\delta\tau^{\prime}_{\rm IR}=\tau/N_3 N_4,$$\delta{\tilde \tau}_{\rm IR}=\tau/N_4$.
This choice results in rather high acceptance rate found 
in Table~\ref{tab:param}. The replay trick\cite{luscher,kennedy} 
is not incorporated. 

For the inversion of the Wilson-Dirac operator during the MD steps 
we implement the same algorithmic techniques as for the run at
$(\kappa_{\rm ud},\kappa_{\rm s})=(0.137\, 810\, 00,0.136\, 400\, 00)$ 
in the previous work\cite{pacscs_nf3}. 
There are three important points to be noted.
First, the initial solution vector is provided by 
the chronological guess with the last 16 solutions\cite{chronological}.
We demand a stringent stopping condition $|Dx-b|/|b|<10^{-14}$ 
to assure the reversibility.
Second, the inversion algorithm is a nested BiCGStab solver consisting
of an inner and an outer solvers. The former plays the role of 
a preconditioner whose calculation is accelerated by single precision
arithmetic with an automatic tolerance control ranging from   
$10^{-3}$ to $10^{-6}$. The latter is implemented with double
precision imposing a stringent tolerance of $10^{-14}$.  
Third, the deflation technique is incorporated in a nested BiCGStab
algorithm: Once the inner solver becomes stagnant during the  
inversion of the Wilson-Dirac operator, the solver algorithm is
automatically replaced by the GCRO-DR (generalized conjugate
residual with implicit inner orthogonalization and deflated restarting)
algorithm\cite{deflation}.
This saves us from the difficulties due to possible small eigenvalues
allowed in the Wilson-type quark action.
We refer to Appendix B in Ref.~\cite{pacscs_nf3} for more details of
the inversion algorithm. 

The strange quark is simulated with the UV-filtered PHMC (UVPHMC) 
algorithm\cite{fastMC,Frezzotti:1997ym,phmc,ishikawa_lat06} 
where the action is UV-filtered\cite{Alexandrou:1999ii} after the
even-odd site preconditioning without domain-decomposition. 
We set the step size as 
$\delta\tau_{\rm s}=\delta\tau^{\prime\prime}_{\rm IR}$
according to our observation  $||F_{\rm s}||\approx ||F_{\rm IR}||$. 
This algorithm is made exact
by correcting the polynomial approximation with the 
global Metropolis test\cite{NoisyMetropolisMB} at the end of each trajectory.
In Table~\ref{tab:param} we find that the choice of $N_{\rm poly}=220$ 
yields 95\% acceptance rate. 
\\  
\section{REWEIGHTING METHOD}
\label{sec:reweighting}

\subsection{Formalism}
\label{subsec:formalizm}

Let us consider evaluating $\la {\cal O}[U](\kappa_{\rm
ud}^\target,\kappa_{\rm s}^\target) \ra_{(\kappa_{\rm
ud}^\target,\kappa_{\rm s}^\target)}$,
which is the expectation value of a physical observable ${\cal O}$
at the target hopping parameters $(\kappa_{\rm ud}^\target,\kappa_{\rm s}^\target)$,
using the configuration samples generated at the original hopping parameters
$(\kappa_{\rm ud},\kappa_{\rm s})$.
We assume that $\rho_{\rm ud}\equiv \kappa_{\rm ud}/\kappa_{\rm ud}^\target\simeq 1$ 
and $\rho_{\rm s}\equiv \kappa_{\rm s}/\kappa_{\rm s}^\target\simeq 1$.
With this assumption,  
the expectation value is rewritten as follows
using the single histogram reweighting method\cite{reweighting}:
\begin{widetext}
\ben
\la {\cal O}[U](\kappa_{\rm ud}^\target,\kappa_{\rm s}^\target) 
\ra_{(\kappa_{\rm ud}^\target,\kappa_{\rm s}^\target)} &=&
\frac{\int {\cal D}U {\cal O}[U](\kappa_{\rm ud}^\target,\kappa_{\rm
s}^\target) \vert \det[D_{\kappa_{\rm ud}^\target}[U]]\vert^2
\det[D_{\kappa_{\rm s}^\target}[U]]{\rm e}^{-S_g[U]}}
{\int {\cal D}U \vert \det[D_{\kappa_{\rm ud}^\target}[U]]\vert^2
\det[D_{\kappa_{\rm s}^\target}[U]]{\rm e}^{-S_g[U]}} \nn\\
&=& 
\frac{
  \int {\cal D}U 
    {\cal O}[U](\kappa_{\rm ud}^\target,\kappa_{\rm s}^\target) 
  \left\vert 
    \det\left[\frac{D_{\kappa_{\rm ud}^\target}[U]}
                   {D_{\kappa_{\rm ud}}[U]} \right] 
  \right\vert^2
    \det\left[\frac{D_{\kappa_{\rm s}^\target}[U]}
                   {D_{\kappa_{\rm s}[U]}}\right]
  \vert \det[D_{\kappa_{\rm ud}}[U]] \vert^2
  \det[D_{\kappa_{\rm s}}[U]]
  {\rm e}^{-S_g[U]}
}
{\int {\cal D}U 
  \left\vert 
    \det\left[\frac{D_{\kappa_{\rm ud}^\target}[U]}
                   {D_{\kappa_{\rm ud}}[U]} \right] 
  \right\vert^2
    \det\left[\frac{D_{\kappa_{\rm s}^\target}[U]}
                   {D_{\kappa_{\rm s}[U]}}\right]
  \vert \det[D_{\kappa_{\rm ud}}[U]] \vert^2
  \det[D_{\kappa_{\rm s}}[U]]
  {\rm e}^{-S_g[U]}
} \nn\\
&=&\frac{\la {\cal O}[U](\kappa_{\rm ud}^\target,\kappa_{\rm s}^\target) 
R_{\rm ud}[U]R_{\rm s}[U]
\ra_{(\kappa_{\rm ud},\kappa_{\rm s})}}
{\la R_{\rm ud}[U]R_{\rm s}[U]\ra_{(\kappa_{\rm ud},\kappa_{\rm s})}},
\label{eq:reweight}
\een
\end{widetext}
where the reweighting factors are defined as
\ben
R_{\rm ud}[U]&=&\left\vert \det\left[\frac{D_{\kappa_{\rm ud}^\target}[U]}
{D_{\kappa_{\rm ud}}[U]} \right] \right\vert^2,\\
R_{\rm s}[U]&=&\det\left[\frac{D_{\kappa_{\rm s}^\target}[U]}
{D_{\kappa_{\rm s}[U]}}\right]
\een
and 
\ben
D_{\kappa_q}[U]=1+\kappa_q (T+M)\;\;\;\; (q={\rm ud},{\rm s})
\een
with $T$ the local clover term including the nonperturbative $\csw$ and
$M$ the hopping matrix.
The above expression (\ref{eq:reweight}) demands us to evaluate the
reweighting factors $R_{\rm ud}[U]$ and $R_{\rm s}[U]$ on each configuration.
For later convenience we define
\ben
W[U](\rho_q)&\equiv &\frac{D_{\kappa_{q}^\target}[U]}
{D_{\kappa_{q}}[U]}
\een
with $\rho_q=\kappa_q/\kappa_{q}^\target$.

\subsection{Evaluation of reweighting factors}
\label{subsec:rewight}

The reweighting factor $R_{\rm ud}[U]$ can be evaluated with a stochastic
method. Introducing a complex bosonic field $\eta$, whose spin and  
color indices are suppressed here, the determinant of $W$ is expressed as
\ben
R_{\rm ud}[U]&=&\left\vert \det\left[W[U](\rho_{\rm ud}) \right]
\right\vert^2\nn\\
&=&\frac{\int {\cal D}\eta^\dagger{\cal D}\eta 
{\rm e}^{-\vert W^{-1}[U](\rho_{\rm ud})\eta \vert^2} }
{\int {\cal D}\eta^\dagger{\cal D}\eta {\rm e}^{-\vert \eta\vert^2}}\nn\\
&=&\frac{\int {\cal D}\eta^\dagger{\cal D}\eta 
{\rm e}^{-\vert W^{-1}[U](\rho_{\rm ud})\eta \vert^2 
+\vert \eta\vert^2 -\vert \eta\vert^2} }
{\int {\cal D}\eta^\dagger{\cal D}\eta {\rm e}^{-\vert \eta\vert^2}}\nn\\
&=&\la {\rm e}^{-\vert W^{-1}[U](\rho_{\rm ud})\eta \vert^2 
+\vert \eta\vert^2 } \ra_\eta,
\een
where $\la \cdots \ra_\eta$ means the expectation value 
with respect to $\eta$.
Given a set of $\eta^{(i)}$ $(i=1,\dots,N_\eta)$ which are random noises
generated according to the Gaussian distribution, the reweighting factor
is evaluated as 
\ben
R_{\rm ud}[U]&=&\lim_{N_\eta \rightarrow\infty} \frac{1}{N_\eta}
\sum_{i=1}^{N_\eta}{\rm e}^{-\vert W^{-1}[U](\rho_{\rm ud})\eta \vert^2 
+\vert \eta\vert^2}.
\een
The ratio $W^{-1}$ is further simplified as follows:
\ben
W^{-1}[U](\rho_{\rm ud})&=&\frac{D_{\kappa_{\rm ud}}[U]}
{D_{\kappa_{\rm ud}^\target}[U]}\nn\\
&=&\rho_{\rm ud}+(1-\rho_{\rm ud})D_{\kappa_{\rm ud}^\target}^{-1}[U]
\een
with the use of $D_{\kappa_{\rm ud}}[U]=\rho_{\rm ud} D_{\kappa_{\rm ud}^\target}[U]+
(1-\rho_{\rm ud})$.
We just need $D_{\kappa_{\rm ud}^\target}^{-1}$ to calculate $W^{-1}$.
 
For the strange quark we assume that 
$\det \left[W[U](\rho_{\rm s})\right]$ is positive.
The corresponding reweighting factor is evaluated as
\ben
R_{\rm s}[U]&=&\det\left[W[U](\rho_{\rm s}) \right]\nn\\
&=&\frac{\int {\cal D}\eta^\dagger{\cal D}\eta 
{\rm e}^{-\vert W^{-1/2}[U](\rho_{\rm s})\eta \vert^2} }
{\int {\cal D}\eta^\dagger{\cal D}\eta {\rm e}^{-\vert \eta\vert^2}}\nn\\
&=&\frac{\int {\cal D}\eta^\dagger{\cal D}\eta 
{\rm e}^{-\vert W^{-1/2}[U](\rho_{\rm s})\eta \vert^2 
+\vert \eta\vert^2 -\vert \eta\vert^2} }
{\int {\cal D}\eta^\dagger{\cal D}\eta {\rm e}^{-\vert \eta\vert^2}}\nn\\
&=&\la {\rm e}^{-\vert W^{-1/2}[U](\rho_{\rm s})\eta \vert^2 
+\vert \eta\vert^2 } \ra_\eta.
\een
With the assumption of $\rho_{\rm s}\simeq 1$ we expect that
$W[U](\rho_{\rm s})$ is so close to the identity matrix 
that its eigenvalues 
are enclosed by a unit circle centered at $(1,0)$ in the complex plane.
In this case we can evaluate $W^{-1/2}[U](\rho_{\rm s})\eta$ 
by the Taylor expansion around identity.

To evaluate the matrix square root $W^{-1/2}[U](\rho_{\rm s})$ 
we first parametrize $W^{-1}[U](\rho_{\rm s})$ as
\ben
W^{-1}[U](\rho_{\rm s})&=&\frac{D_{\kappa_{\rm s}}[U]}
{D_{\kappa_{\rm s}^\target}[U]}\nn\\
&=&\rho_{\rm s}+(1-\rho_{\rm s})D^{-1}_{\kappa_{\rm s}^\target}[U]\nn\\
&=&1-(1-\rho_{\rm s})\left(1-D^{-1}_{\kappa_{\rm s}^\target}[U]\right)\nn\\
&=&1-X[U](\rho_{\rm s})
\een
where $\vert 1-\rho_{\rm s}\vert \ll 1$ and $\| X[U](\rho_{\rm s}) \|<1$.
We employ the recursive expression for the Taylor expansion of  
$W^{-1/2}[U](\rho_{\rm s})\eta$\cite{phmc}:
\begin{widetext}
\ben
W^{-1/2}\eta &=&\sum_{j=0}^{N} c_j X^j\eta \nn\\
&=&c_0\left[\eta+\frac{c_1}{c_0}X
\left[\eta+\frac{c_2}{c_1}X 
\left[\eta+\frac{c_3}{c_2}X
\left[\cdots\left[\eta+\frac{c_{N-1}}{c_{N-2}}X
\left[\eta+\frac{c_{N}}{c_{N-1}}X\eta \right] \right]\right] 
\right]\right]\right],
\een
\end{widetext}
where the argument $[U](\rho_{\rm s})$ for the matrices is suppressed.
The coefficients are given by
\ben
\frac{c_j}{c_{j-1}}=1-\frac{3}{2j}
\een
with $c_0=1$.
The advantage of the recursive procedure is to reduce the round-off
errors in the summation from the lower-order to the higher-order
contributions in the Taylor expansion.
The truncation error  and the order of the Taylor expansion $N$
are monitored and controlled during the simulation 
by explicitly evaluating 
the residual $r=||(W^{-{1/2}}W^{-{1/2}}-W^{-1} )\eta||//||\eta||$.
We enforce the condition $r < 10^{-14}$ for $N$.

To reduce the fluctuations in the stochastic evaluation of 
$R_{\rm ud}[U]$ and $R_{\rm s}[U]$ we employ the determinant 
breakup technique\cite{hasenfratz,jung}.
The interval between $\kappa_q$ and $\kappa_q^\target$ is divided into
$N_B$ subintervals: $\{ \kappa_q$, $\kappa_q+\Delta_q$, $\dots$,
$\kappa_q+(N_B-1)\Delta_q$, $\kappa_q^\target \}$ with 
$\Delta_q=(\kappa_q^\target-\kappa_q )/N_B$.
Thus the determinant of $W[U](\rho_{q})$ is broken up as
\begin{widetext}
\ben
\det \left[W[U](\rho_{q})\right]&=&
\det \left[W[U]\left(\frac{\kappa_q}{\kappa_q+\Delta_q}\right)\right]\cdot
\det \left[W[U]\left(\frac{\kappa_q+\Delta_q}{\kappa_q+2\Delta_q}\right)\right]\cdots
\det \left[W[U]\left(\frac{\kappa_q+(N_B-1)\Delta_q}{\kappa_q^\target}\right)\right],\nn
\label{eq:reweightbreakuped}\\
\een
\end{widetext}
where each determinant in the right hand side is evaluated with
an independent noise set of $\eta$. 
For strange quark reweighting, $W^{-1}$ are simply replaced by $W^{-1/2}$ in
Eq.~(\ref{eq:reweightbreakuped}).

\subsection{Parameters and results for reweighting factors}
\label{subsec:result_rw}

Our choice of the target hopping parameters are $(\kappa_{\rm
ud}^\target,\kappa_{\rm s}^\target)=(0.137\, 796\, 25,0.136\, 633\, 75)$. The subintervals 
for the determinant breakup are $\Delta_{\rm ud}=(0.137\, 796\, 25-0.137\, 785\, 00)/N_B$ 
with $N_B=3$ for the ud quark and $\Delta_{\rm s}=(0.136\, 633\, 75-0.136\, 600\, 00)/N_B$ 
with $N_B=3$ for
the s quark. Each piece of the divided determinant is evaluated 
stochastically employing 10 sets of $\eta$ 
at every 100 trajectories (25 MD time units).
The order of Taylor expansion $N$ was mostly 5 for each of
the strange quark reweighting break up.

\begin{figure*}[h]
\vspace{13mm}
\begin{center}
\includegraphics[width=90mm,angle=0]{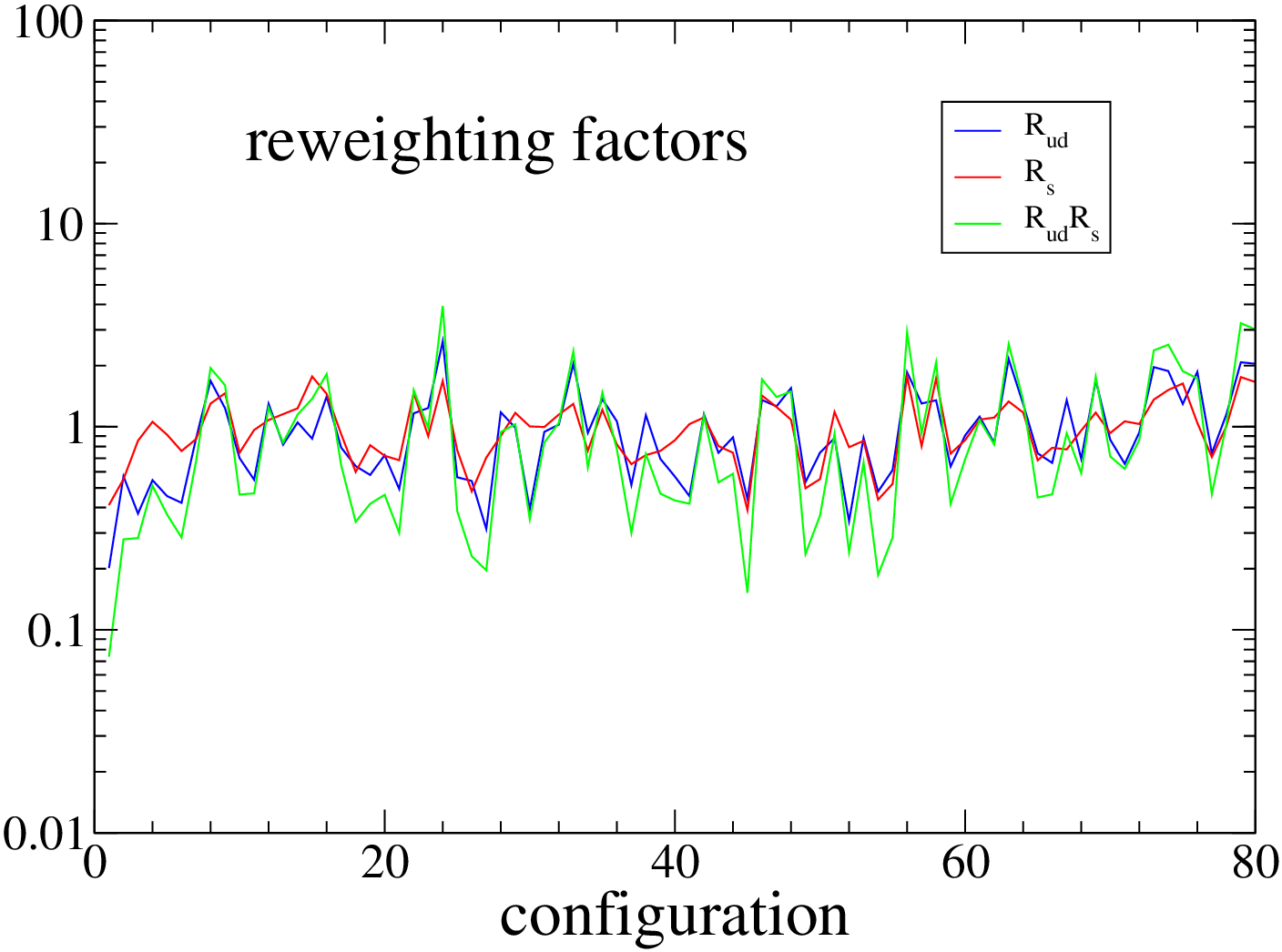}
\end{center}
\vspace{-.5cm}
\caption{Configuration dependence of reweighting factors from 
$(\kappa_{\rm ud},\kappa_{\rm s})=(0.137\,785\,00,0.136\,600\,00)$ to 
$(\kappa_{\rm ud}^\target,\kappa_{\rm s}^\target)=(0.137\,796\,25,0.136\,633\,75)$.}
\label{fig:rwfactor_c}
\end{figure*}

\begin{figure*}[h]
\vspace{13mm}
\begin{center}
\includegraphics[width=90mm,angle=0]{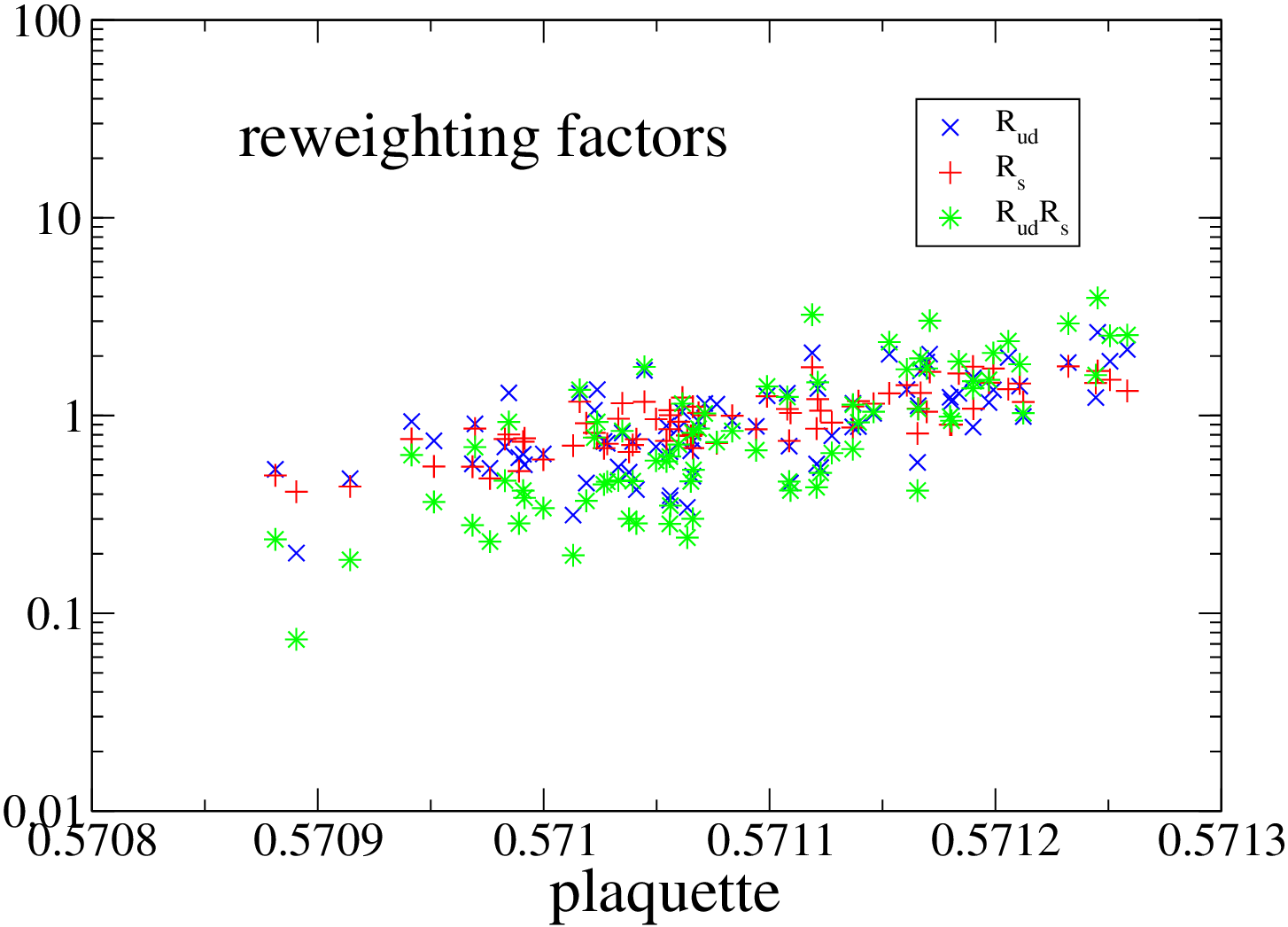}
\end{center}
\vspace{-.5cm}
\caption{Reweighting factors from 
$(\kappa_{\rm ud},\kappa_{\rm s})=(0.137\,785\,00,0.136\,600\,00)$ to 
$(\kappa_{\rm ud}^\target,\kappa_{\rm s}^\target)=(0.137\,796\,25,0.136\,633\,75)$
as a function of plaquette value.}
\label{fig:rwfactor_p}
\end{figure*}

\begin{figure*}[h]
\vspace{13mm}
\begin{center}
\includegraphics[width=85mm,angle=0]{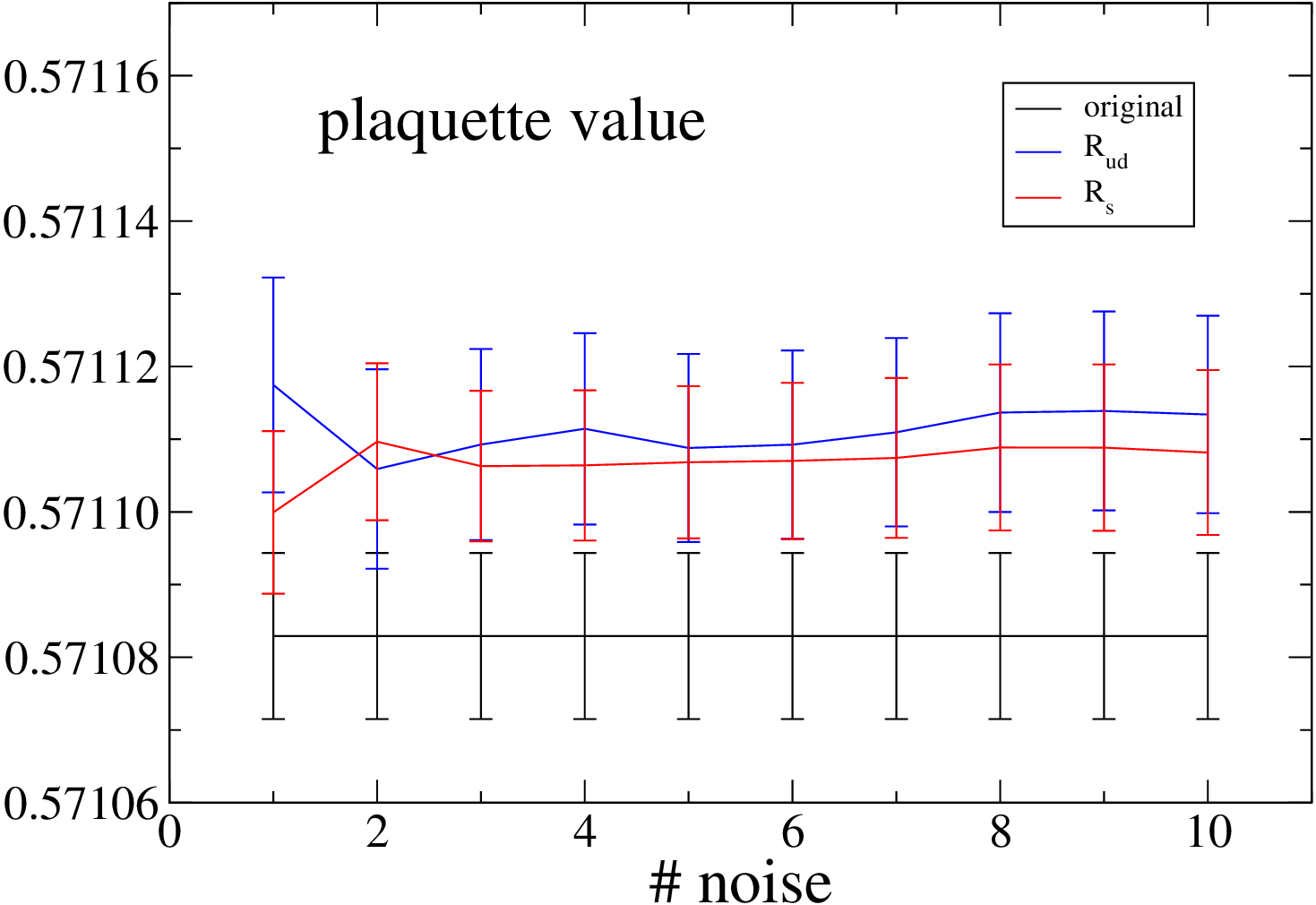}  
\end{center}
\vspace{-.5cm}
\caption{Reweighted plaquette values with $R_{\rm ud}$ and $R_{\rm s}$ as a function of the number of noise.}
\label{fig:plaq_rw}
\end{figure*}

Figure~\ref{fig:rwfactor_c} shows configuration dependence of 
the reweighting factors from 
$(\kappa_{\rm ud},\kappa_{\rm s})=(0.137\, 785\, 00,0.136\, 600\, 00)$ 
to $(\kappa_{\rm
ud}^\target,\kappa_{\rm s}^\target)=(0.137\, 796\, 25,0.136\, 633\, 75)$
which are normalized as $\la R_{\rm ud,s}\ra=1$ and 
$\la R_{\rm ud}R_{\rm s}\ra=1$.
The fluctuations of $R_{\rm ud}$ and $R_{\rm
s}$ are within a factor of 10. Their product has slightly amplified
fluctuations. 
In Fig.~\ref{fig:rwfactor_p} we plot the reweighting factors
as a function of the plaquette value on each configuration. 
An important observation is a clear correlation between the reweighting factors and the plaquette value: The former increases as the latter becomes larger.
Thanks to this correlation the
distribution of the plaquette value at $(\kappa_{\rm
ud}^\target,\kappa_{\rm s}^\target)=(0.137\, 796\, 25,0.136\, 633\, 75)$ is moved in the
positive direction. 
This is the expected behavior, because the target hopping parameters are larger than the original ones.
The situation is quantitatively illustrated in Fig.~\ref{fig:plaq_rw},
where the reweighted plaquette values with $R_{\rm ud}$ and $R_{\rm s}$ 
are individually plotted as a function of 
the number of noise. The results look converged once the number of noise goes
beyond four. 

\begin{figure*}[h]
\vspace{13mm}
\begin{center}
\includegraphics[width=85mm,angle=0]{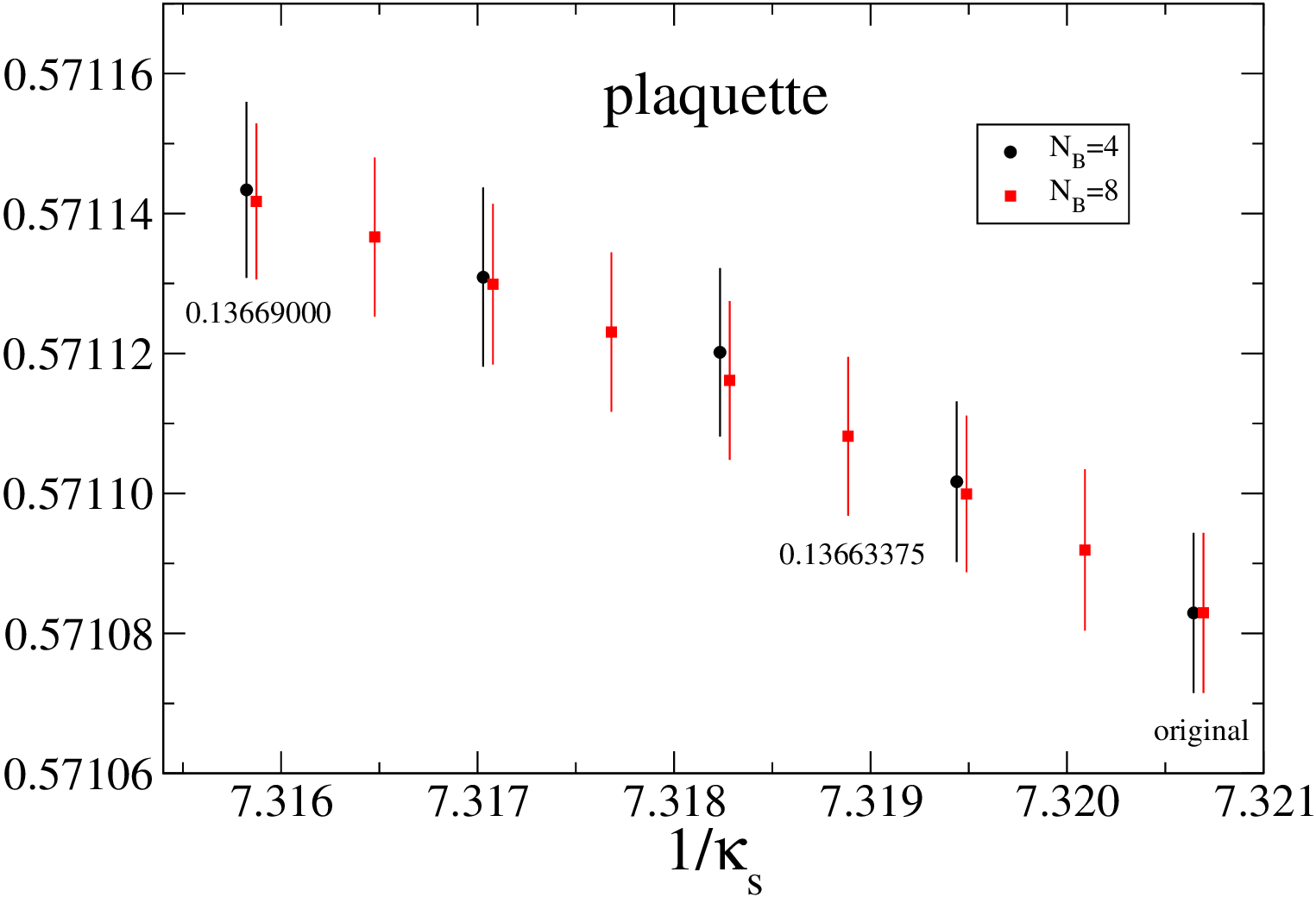}  
\end{center}
\vspace{-.5cm}
\caption{Reweighted plaquette values with $R_{\rm s}$ as a function of
target value of $\kappa_{\rm s}$. 
Interval from $\kappa_{\rm s}=0.136\,600\,00$ to 0.136\,690\,00
is divided by $N_B=4$ (black) and 8 (red).}
\label{fig:test_plaq_s}
\end{figure*}

\begin{figure*}[h]
\vspace{13mm}
\begin{center}
\includegraphics[width=85mm,angle=0]{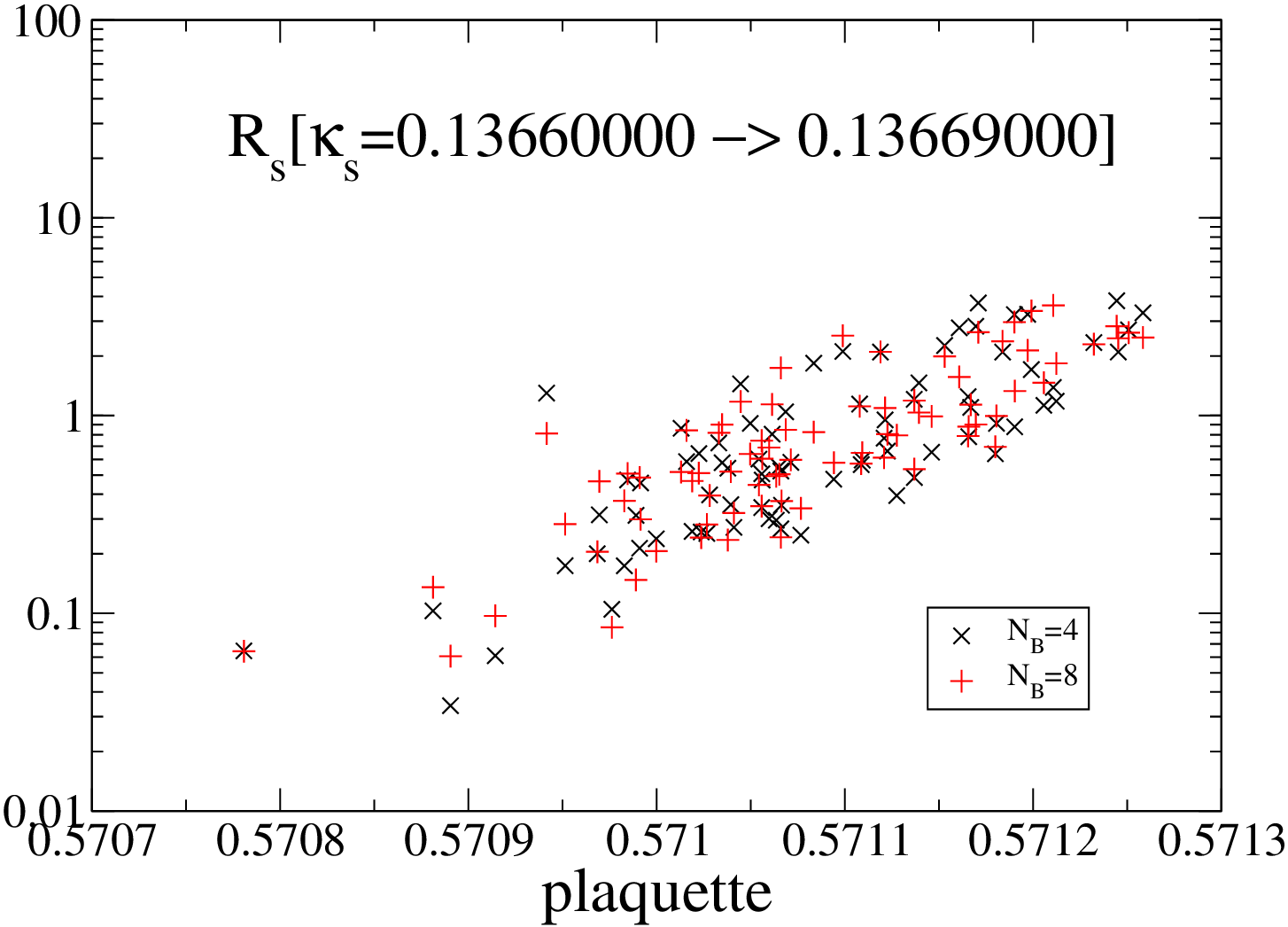}  
\end{center}
\vspace{-.5cm}
\caption{Reweighting factor $R_{\rm s}$ 
from $\kappa_{\rm s}=0.136\,600\,00$ to 0.136\,690\,00
with $N_B=4$ (black) and 8 (red) as a function of plaquette value.}
\label{fig:test_rw_p_s}
\end{figure*}

\begin{figure*}[h]
\vspace{13mm}
\begin{center}
\includegraphics[width=85mm,angle=0]{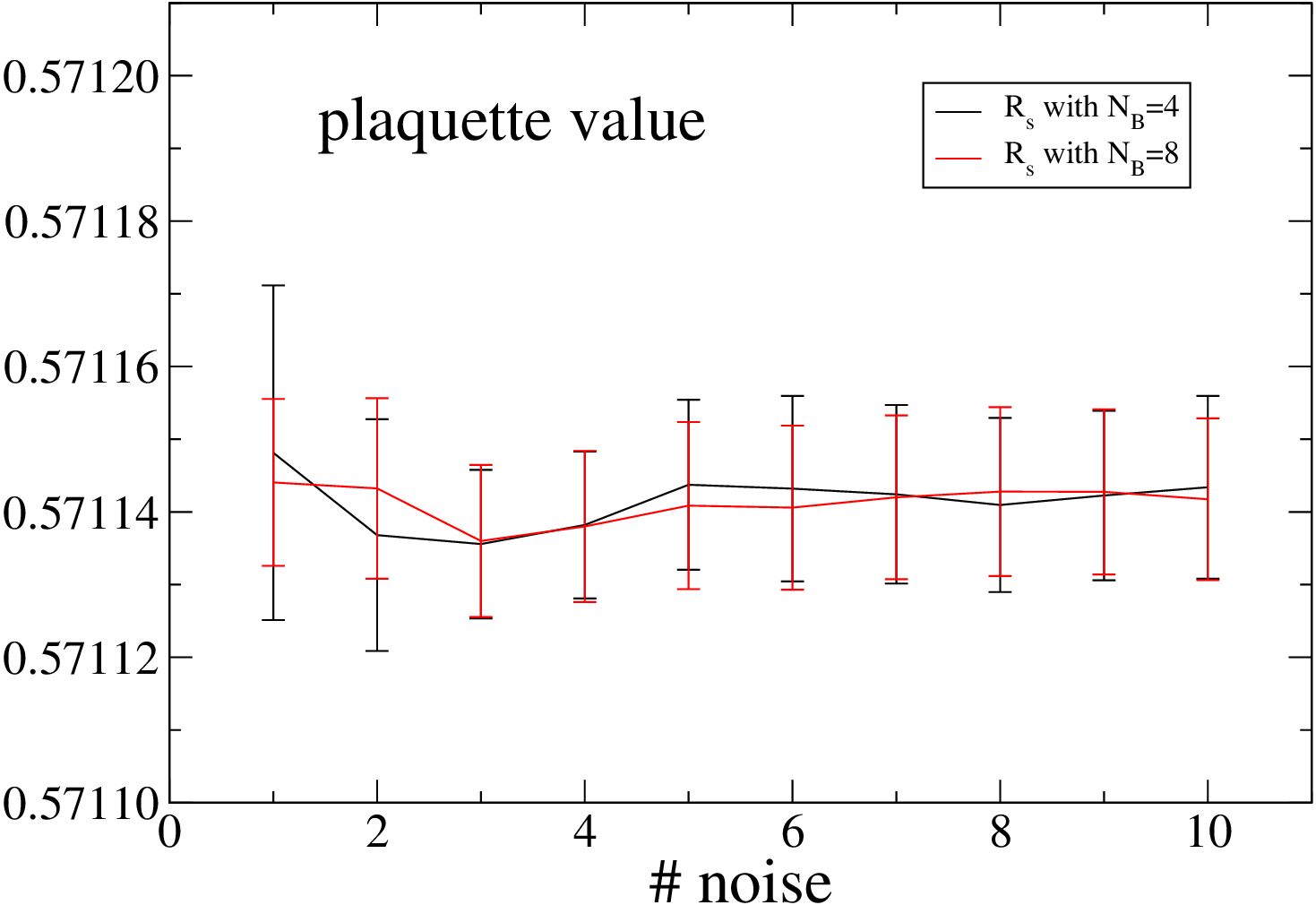}  
\end{center}
\vspace{-.5cm}
\caption{Reweighted plaquette value with $R_{\rm s}$ 
from $\kappa_{\rm s}=0.136\,600\,00$ to 0.136\,690\,00 as a function of 
the number of noise. Interval 
is divided by $N_B=4$ (black) and 8 (red).}
\label{fig:test_plaq_rw_s}
\end{figure*}

Since the formula of Eq.~\ref{eq:reweight} is the identity,
the reweighting procedure is always assured if we have infinite statistics.
In case of finite statistics in practical simulations, however, 
we should be concerned with 
the possible situation that the original 
and the target points are far away such that the distributions of observable
fail to overlap each other.
This problematic case could be detected by 
monitoring the behavior of the expectation value for the observable
as the reweighting parameters are monotonically moved from the original
point: The expectation value of the observable stops varying with
diminishing error bar.
To check the reliability of our reweighting procedure
we have investigated the behavior of the expectation value of the 
plaquette against the reweighting with respect to the strange quark from
$\kappa_{\rm s}=0.136\, 600\, 00$ to $\kappa_{\rm s}=0.136\, 690\, 00$ 
with $N_B=4$ and 8, the latter of which yields the same amount of
breakup interval as 
$\Delta_{\rm s}=(0.136\, 633\, 75-0.136\, 600\, 00)/3=0.000\, 011\, 25$ in our choice.
Since the plaquette value has much narrower distribution than the
hadron propagators at each time slice, this is a stringent test
to check the overlap of the distributions of the observable
at the original and the target points.  
Figure~\ref{fig:test_plaq_s} shows the behavior of the reweighted plaquette 
value evaluated with 10 noise sources 
as a function of the reweighting parameter $\kappa_{\rm s}$.
We do not observe any sign that the reweighted plaquette value 
stagnates at some point:
It shows almost linear behavior with constant magnitude of error
up to $\kappa_{\rm s}=0.136\, 690\, 00$ which
is far beyond the physical point of $\kappa_{\rm s}=0.136\, 633\, 75$.
Furthermore $N_B=4$ and 8 cases give consistent results.
In Fig.~\ref{fig:test_rw_p_s} we plot the reweighting factor  
$R_{\rm s}$ from $\kappa_{\rm s}=0.136\, 600\, 00$ 
to $\kappa_{\rm s}=0.136\, 690\, 00$ with $N_B=4$ and 8
as a function of the plaquette value on each configuration,
which is normalized as $\la R_{\rm s}\ra=1$.
Both cases show quite similar distributions, which confirm that
our choice of breakup interval 
$\Delta_{\rm s}=0.000\, 011\, 25$ is sufficiently small.
In Fig.~\ref{fig:test_plaq_rw_s} we also present 
the reweighted plaquette value with $R_{\rm s}$
as a function of the number of noise.
The results with $N_B=4$ and 8 become fairly consistent once 
we employ more than two noise sources. 
We have repeated the same analyses
for the reweighting with respect to the ud-down quark from
$\kappa_{\rm s}=0.137\, 785\, 00$ to $\kappa_{\rm s}=0.137\, 800\, 00$ 
with $N_B=2$ and 4.
The same conclusion is obtained as in the strange quark case.
This is easily expected from similar behaviors 
for $R_{\rm ud}$ and $R_{\rm s}$ 
found in Figs.~\ref{fig:rwfactor_c}, \ref{fig:rwfactor_p} 
and \ref{fig:plaq_rw}.

\section{HADRONIC OBSERVABLES}
\label{sec:hadron}

\subsection{Hadron masses, quark masses and decay constants at
  simulation point}
\label{subsec:hmass_o}

We measure the meson and the baryon correlators employing  
appropriate operators.
The general form of the meson operators is expressed as
\ben
M_\Gamma^{fg}(x)={\bar q}_f(x)\Gamma q_g(x),
\een
where $f$ and $g$ denote quark flavors and $\Gamma$ are 16
Dirac matrices $\Gamma={\rm I}$, $\gamma_5$, $\gamma_\mu$, 
$i\gamma_\mu\gamma_5$ and  $i[\gamma_\mu,\gamma_\nu]/2$ 
$(\mu,\nu=1,2,3,4)$.
The octet baryon operators are given by 
\ben
{\cal O}^{fgh}_\alpha(x)=\epsilon^{abc}((q_f^a(x))^T C\gamma_5 q_g^b(x))
q_{h\alpha}^c(x),
\een 
where $a,b,c$ are color indices, $C=\gamma_4\gamma_2$ is the 
charge conjugation matrix and $\alpha=1,2$ labels the $z$-component
of the spin 1/2.
The $\Sigma$- and $\Lambda$-like octet baryons are distinguished by
the flavor structures:
\ben
\Sigma{\rm -like}\;\; &:& \;\; -\frac{{\cal O}^{[fh]g}+{\cal O}^{[gh]f}}{\sqrt{2}},\\
\Lambda{\rm -like}\;\; &:& \;\; \frac{{\cal O}^{[fh]g}-{\cal O}^{[gh]f}-2{\cal O}^{[fg]h}}
{\sqrt{6}},
\een
where $O^{[fg]h}={\cal O}^{fgh}-{\cal O}^{gfh}$.
We define the decuplet  baryon operators for the four $z$-components of the
spin 3/2 as
\ben
D^{fgh}_{3/2}(x)&=&\epsilon^{abc}((q_f^a(x))^T C\Gamma_+
q_g^b(x))q_{h1}^c(x),\\
D^{fgh}_{1/2}(x)&=&
\epsilon^{abc}[((q_f^a(x))^T C\Gamma_0q_g^b(x))q_{h1}^c(x)\nn\\
&&-((q_f^a(x))^T C\Gamma_+q_g^b(x))q_{h2}^c(x)]/3,\\
D^{fgh}_{-1/2}(x)&=&
\epsilon^{abc}[((q_f^a(x))^T C\Gamma_0q_g^b(x))q_{h2}^c(x)\nn\\
&&-((q_f^a(x))^T C\Gamma_-q_g^b(x))q_{h1}^c(x)]/3,\\
D^{fgh}_{-3/2}(x)&=&\epsilon^{abc}((q_f^a(x))^T C\Gamma_-
q_g^b(x))q_{h2}^c(x),
\een
where $\Gamma_{\pm}=(\gamma_1\mp\gamma_2)/2$, $\Gamma_0=\gamma_3$ and
the flavor structures should be symmetrized.

The meson and the baryon correlators are calculated 
with point and smeared sources and a local sink.
The smeared source is constructed with an exponential smearing function 
$\Psi(|{\vec x}|)=A_q\exp(-B_q|{\vec x}|)$ $(q={\rm ud,s})$ where
$\Psi(0)=1$ for the ud and s quark propagators. 
Employing a couple of thermalized configurations
we adjust the parameters such that the pseudoscalar meson effective masses
reach a plateau as soon as possible.
Our choice is $A_{\rm ud}=1.2, B_{\rm ud}=0.07$ and 
$A_{\rm s}=1.2, B_{\rm s}=0.18$.

To reduce the statistical error of the zero momentum hadron correlators 
we employ two methods.
One is the choice of four source points 
at $(x_0,y_0,z_0,t_0)$=$(17,17,17,1)$, $(1,1,1,9)$,
$(25,25,25,17)$, and $(9,9,9,25)$.
The other is the use of possible spin states:
three polarization states for the vector meson and two (four) 
spin states for the octet (decuplet) baryons. 
The correlators with different sources and spin states 
are averaged on each configuration  before the jackknife analysis. 

\begin{figure*}[h]
\vspace{13mm}
\begin{center}
\begin{tabular}{cc}
\includegraphics[width=85mm,angle=0]{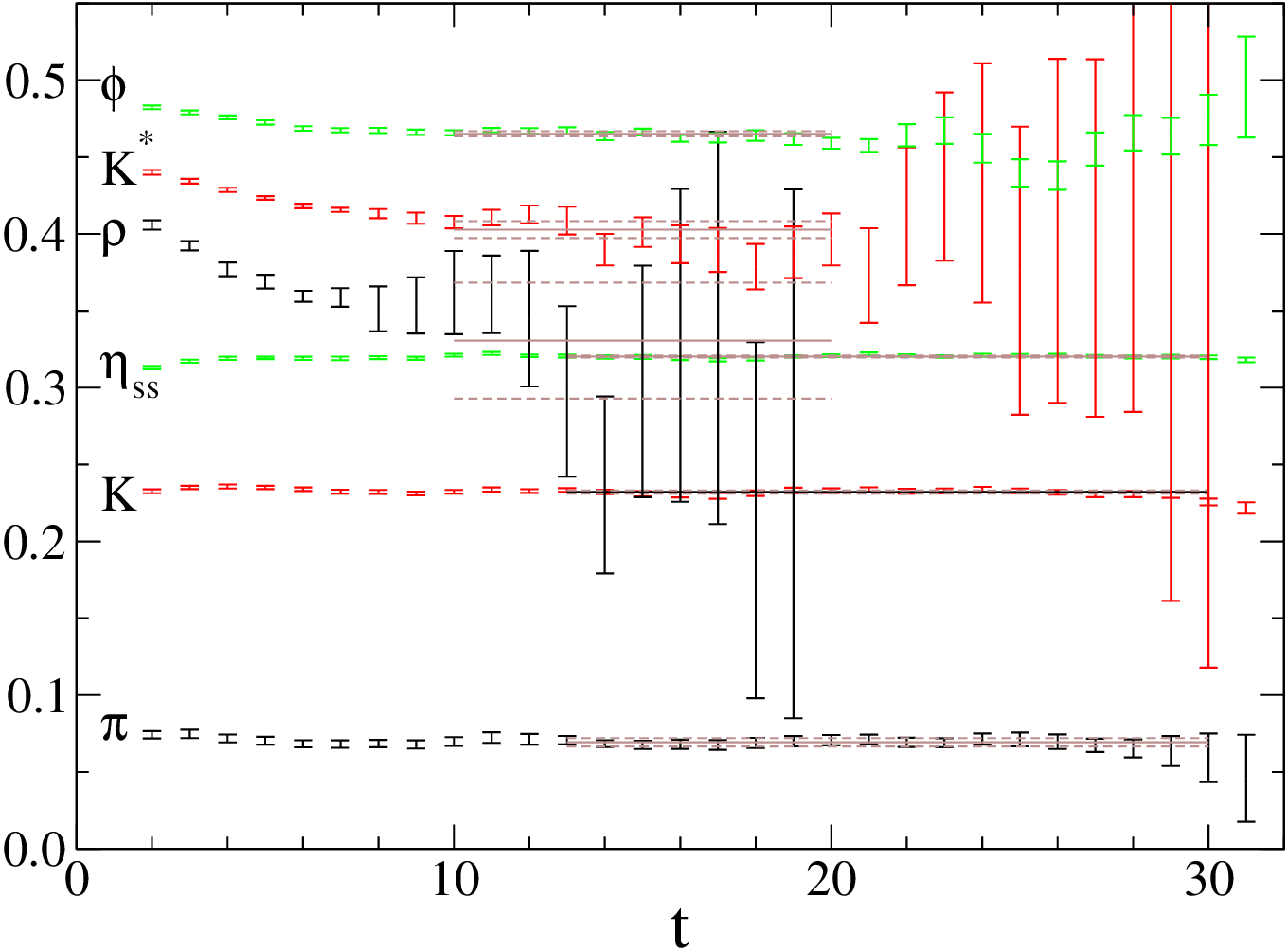} 
\hspace*{2mm}&\hspace*{2mm}
\includegraphics[width=85mm,angle=0]{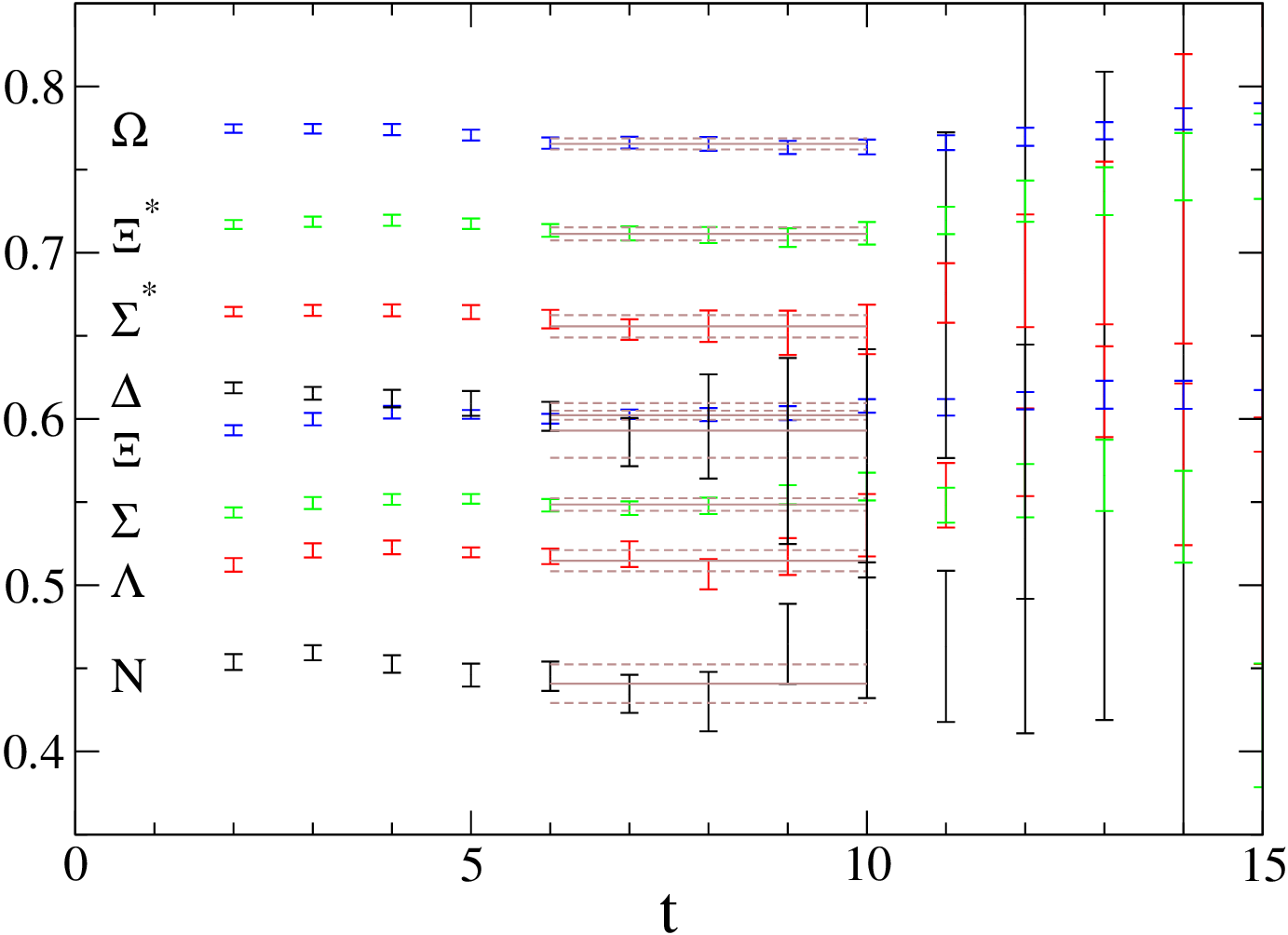}
\end{tabular}
\end{center}
\vspace{-.5cm}
\caption{Effective masses for the mesons (left) and the baryons (right) 
at the simulation point of 
$(\kappa_{\rm ud},\kappa_{\rm s})=(0.137\,785\,00,0.136\,600\,00)$.
Horizontal bars represent the fit results with 1 standard 
deviation error band.}
\label{fig:m_eff_o}
\end{figure*}
\begin{figure*}[h]
\vspace{13mm}
\begin{center}
\includegraphics[width=85mm,angle=0]{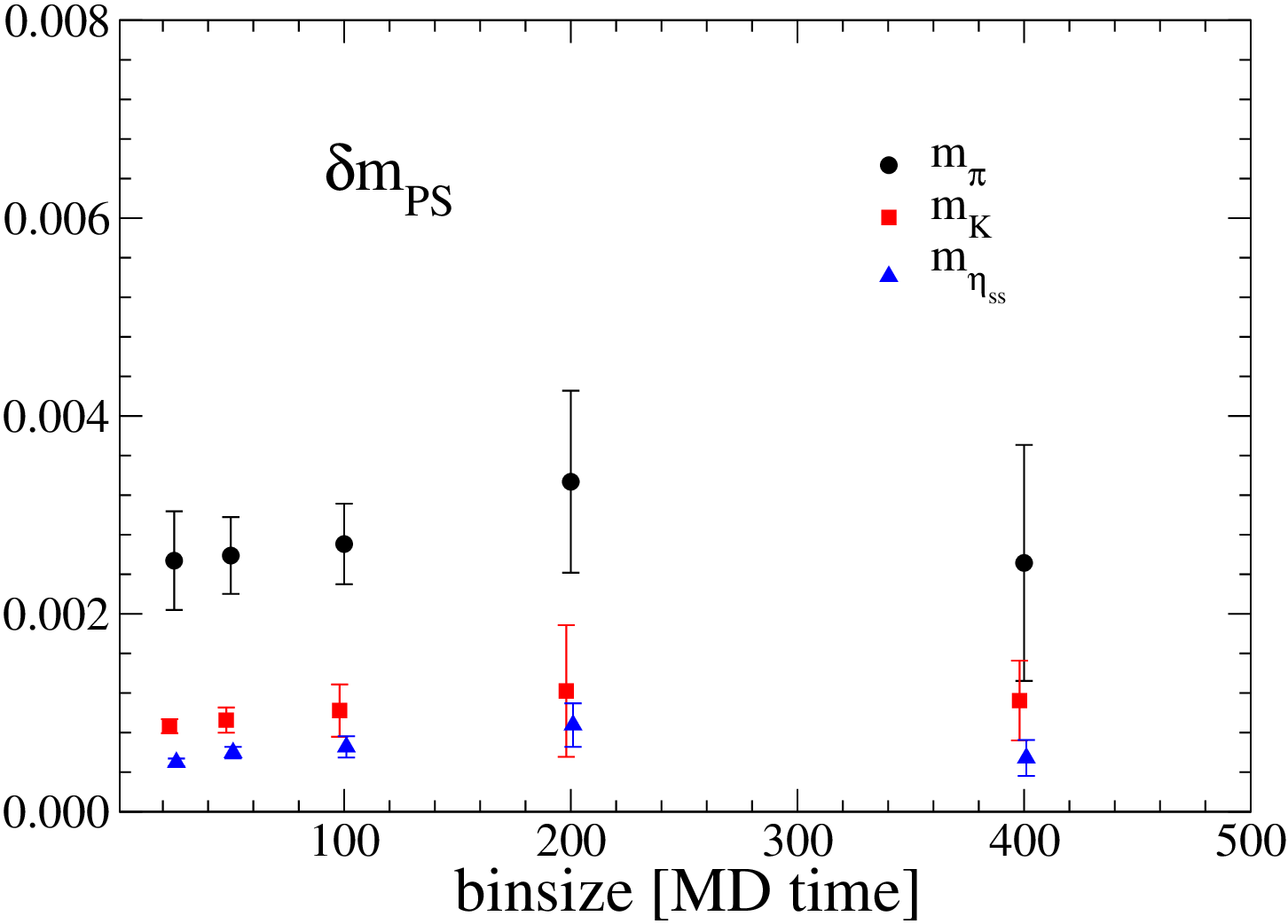}  
\end{center}
\vspace{-.5cm}
\caption{Binsize dependence of the magnitude of error for
the pseudoscalar meson masses.}
\label{fig:binerr_ps}
\end{figure*}

Figure~\ref{fig:m_eff_o} shows effective mass plots  
for the meson and baryon propagators with the smeared
source, where we assume a single
hyperbolic cosine function for the former 
and a single exponential form for the latter.
We observe good plateaux starting at small values of $t$, showing that  
the excited state contributions are suppressed. 
The hadron masses are extracted by uncorrelated $\chi^2$ fit to the
propagators, since we find instabilities in correlated fit
using covariance matrix.  
The horizontal bars in Fig.~\ref{fig:m_eff_o} represent
the fit ranges,
which are $[t_{\rm min},t_{\rm max}]=[13,30]$ 
for the pseudoscalar mesons, $[10,20]$ for the
vector mesons and $[6,10]$ for the baryons, 
and the resulting hadron masses 
with 1 standard deviation error band.
The numerical values are summarized in Table~\ref{tab:hmass_lat}. 
The statistical errors are estimated with the jackknife
method. 
In Fig.~\ref{fig:binerr_ps} we plot the binsize dependence of the
error for the pseudoscalar meson masses.
The magnitude of error shows flat behaviors against
the binsize within the error bars. 
Since similar binsize dependences are found for other
particle types, we employ a binsize of 100 MD time
(4 gauge configurations) for the jackknife analysis.
As a cross check we also carry out the bootstrap error estimation
with 5000 samples. For all the physical quantities at the original and the
target points the bootstrap samples show clear normal distribution 
and the error estimates agree with those of the jackknife method
within 10\%.


For the quark masses and the decay constants we have accomplished
an important improvement since the previous publication\cite{pacscs_nf3}:
a nonperturbative determination of renormalization factors 
based on the Sch{\"o}dinger functional scheme\cite{npr_alpha,tani_lat09,npr_zm}.
The bare quantities are calculated with the same method 
as in Ref.~\cite{pacscs_nf3}.  
 
The bare quark mass is defined by the axial vector
Ward-Takahashi identity (AWI): 
\ben
{\bar m}^{\rm AWI}_f+ {\bar m}^{\rm AWI}_g=\frac{\langle 0 |\nabla_4
  A_4^{\rm imp} |{\rm PS} \rangle}{\langle 0| P | PS\rangle},
\een
where $P$ is the pseudoscalar operator and $|{\rm PS}\rangle$ denotes
the pseudoscalar meson state at rest consisting of $f$ and $g$
$(f,g={\rm ud},{\rm s})$ valence quarks.
The axial vector current is nonperturbatively 
$O(a)$-improved as $A_4^{\rm imp}=A_4+c_A{\bar \nabla}_4 P$ 
with ${\bar \nabla}_4$ the symmetric lattice derivative and 
$c_A=-0.038\, 761\, 06$\cite{ca}.
The ratio of the matrix elements is evaluated by
\ben
{\bar m}^{\rm AWI}_f+{\bar m}^{\rm AWI}_g&=&m_{\rm PS}
\left\vert \frac{C_A^s}{C_P^s}\right\vert,
\een
where $m_{\rm PS}$, $C_A^s$ and $C_P^s$ are extracted from
a simultaneous $\chi^2$ fit to 
\ben
\langle A_4^{\rm imp}(t) P^s(0)\rangle=2C_A^s
\frac{\sinh(-m_{\rm PS}(t-T/2))}{\exp(m_{\rm PS}T/2)}
\label{eq:a4p_s}
\een
and
\ben
\langle P(t)P^s(0)\rangle=2C_P^s
\frac{\cosh(-m_{\rm PS}(t-T/2))}{\exp(m_{\rm PS}T/2)}
\label{eq:pp_s}
\een
with $P^s$ the smeared pseudoscalar operator and
$T=64$ the temporal extent of the lattice.
The fit ranges are chosen to be $[t_{\rm min},t_{\rm max}]=[13,25]$ 
for the former and $[13,30]$ for the latter.
The renormalized quark mass in the continuum ${\overline{\rm MS}}$
scheme is defined as 
\ben
m^{\overline{\rm MS}}_f&=&Z_m^\msbar{\bar m}^{\rm AWI}_f,
\een
where $Z_m^\msbar=Z_A/Z_P$ is nonperturbatively determined in the Schr{\"o}dinger
functional scheme.  
In Table~\ref{tab:qmass} we present the results for $m^{\overline{\rm
MS}}_{\rm ud}$ and $m^{\overline{\rm MS}}_{\rm s}$ renormalized at $\mu=2$
GeV together with the corresponding
bare quark masses ${\bar m}^{\rm AWI}_f$ and ${\bar m}^{\rm AWI}_{\rm s}$.
We use $Z_m^\msbar($2{\rm GeV})=1.441(15)\cite{npr_zm}. 
The statistical errors are estimated
by the jackknife analysis with the choice of 
the same binsize as for the hadron masses.

The bare pseudoscalar meson decay constant defined by
\ben
\sqrt{2\kappa_f}\sqrt{2\kappa_g}
\left\vert\langle 0|A_4^{\rm imp}|{\rm PS}\rangle\right\vert 
=f^{\rm bare}_{\rm PS}m_{\rm PS}.
\een
is evaluated from the following formula:
\ben
f^{\rm bare}_{\rm PS}&=&\sqrt{2\kappa_f}\sqrt{2\kappa_g}
\left\vert\frac{C_A^s}{C_P^s}\right\vert
\sqrt{\frac{2\left\vert C_P^l\right\vert}{m_{\rm PS}}}.
\een
We extract $m_{\rm PS}$,  $C_A^s$ , $C_{P}^s$ and $C_P^l$
from a simultaneous fit of Eqs.~(\ref{eq:a4p_s}), (\ref{eq:pp_s}) and
\ben
\langle P(t)P^l(0)\rangle=2C_P^l
\frac{\cosh(-m_{\rm PS}(t-T/2))}{\exp(m_{\rm PS}T/2)}
\label{eq:pp_l}
\een
with $P^l$ the local pseudoscalar operator.
The fit ranges are $[13,25]$, $[13,30]$ and $[15,25]$, respectively.
The renormalization is given by
\ben
f_{\rm PS}&=&Z_A f^{\rm bare}_{\rm PS},
\een
with $Z_A=0.8563(52)$\cite{npr_zm} the nonperturbative renormalization factor in the
Schr{\"o}dinger functional scheme.
In Table~\ref{tab:qmass} we list 
the results for $f_{\rm PS}$ and $f^{\rm bare}_{\rm PS}$ 
with the statistical errors evaluated 
in the same manner as for the quark masses.

\subsection{Hadron masses, quark masses and decay constants at target point}
\label{subsec:hmass_t}
 
\begin{figure*}[h]
\vspace{13mm}
\begin{center}
\begin{tabular}{cc}
\includegraphics[width=85mm,angle=0]{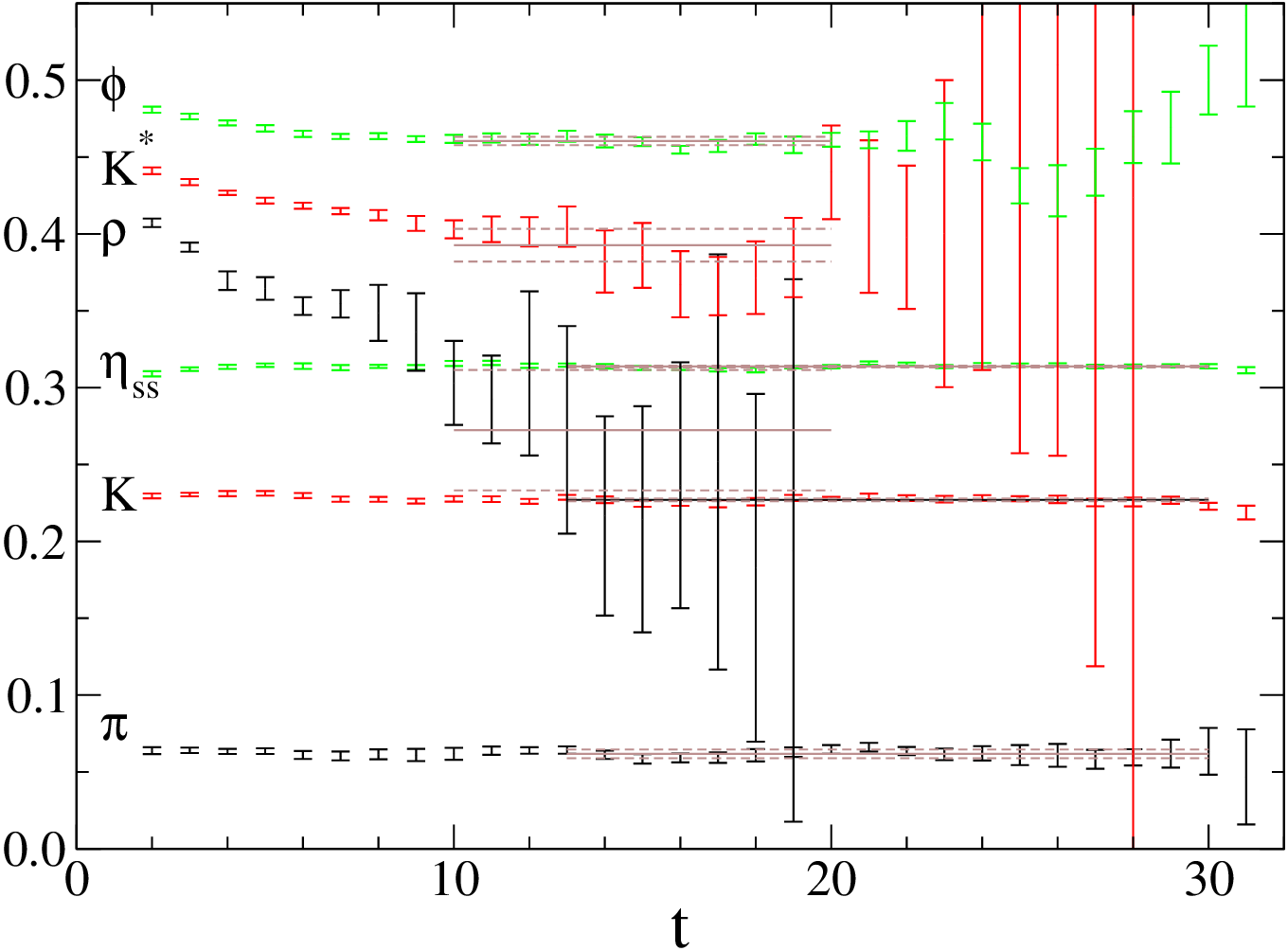} 
\hspace*{2mm}&\hspace*{2mm}
\includegraphics[width=85mm,angle=0]{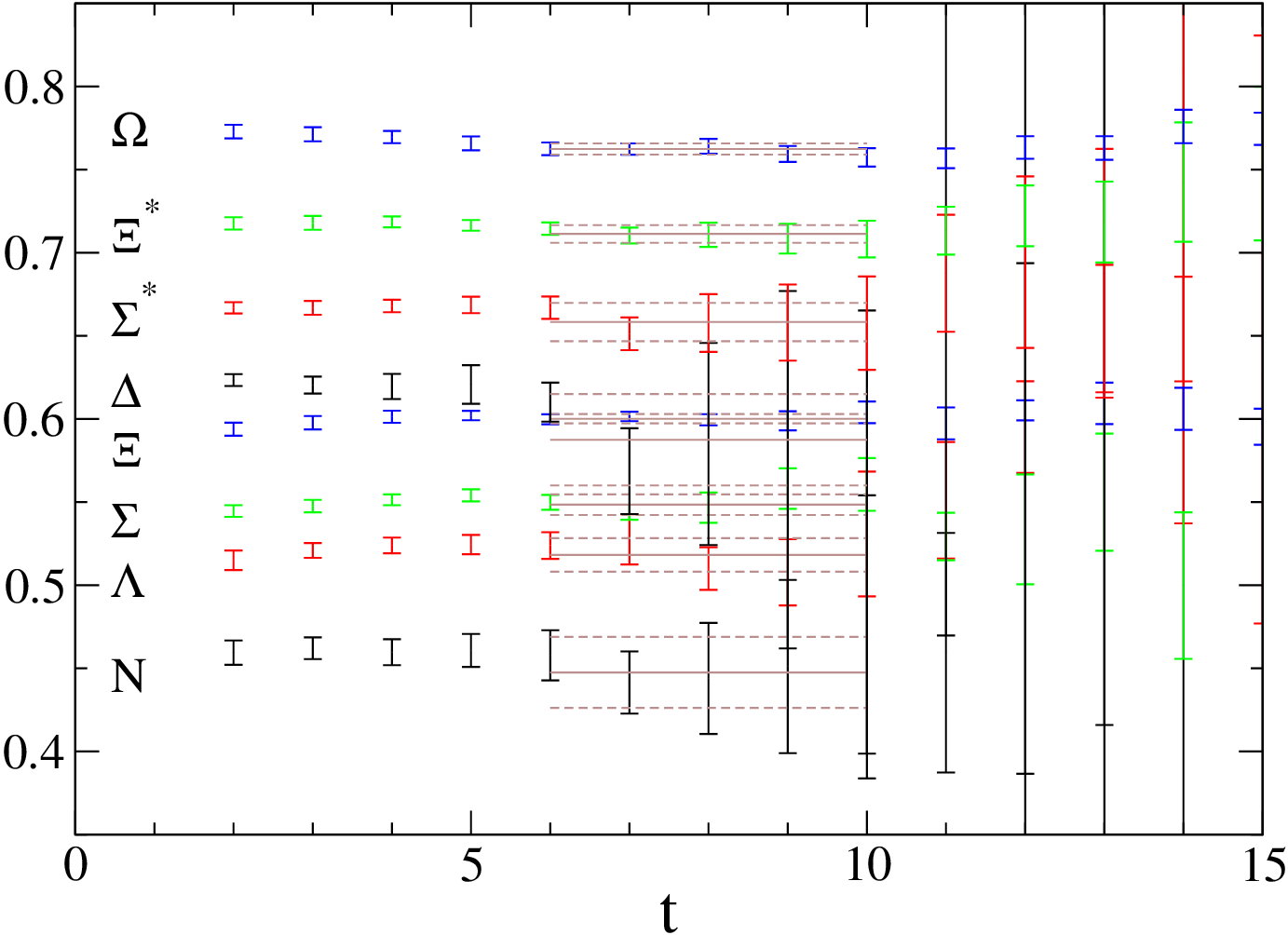}
\end{tabular}
\end{center}
\vspace{-.5cm}
\caption{Effective masses for the mesons (left) and the baryons (right) 
at the target point of 
$(\kappa_{\rm ud}^\target,\kappa_{\rm s}^\target)=(0.137\,796\,25,0.136\,633\,75)$.
Horizontal bars represent the fit results with 1 standard 
deviation error band.}
\label{fig:m_eff_t}
\end{figure*}

In Fig.~\ref{fig:m_eff_t} we present the effective masses for the reweighted
meson and baryon propagators with the smeared source.  
Comparing with the original case in Fig.~\ref{fig:m_eff_o} 
the error bars are slightly enlarged by the reweighting procedure. 
We apply the uncorrelated $\chi^2$ fit to the reweighted hadron
propagators at the target point choosing the same fit ranges and
jackknife binsize as in the simulation point. The results are summarized 
in Tables~\ref{tab:hmass_lat} and \ref{tab:hmass_ph}, where we also present 
the previous results obtained by the chiral extrapolation method 
in Ref.~\cite{pacscs_nf3} for comparison.

\begin{figure*}[h]
\vspace{13mm}
\begin{center}
\begin{tabular}{cc}
\includegraphics[width=85mm,angle=0]{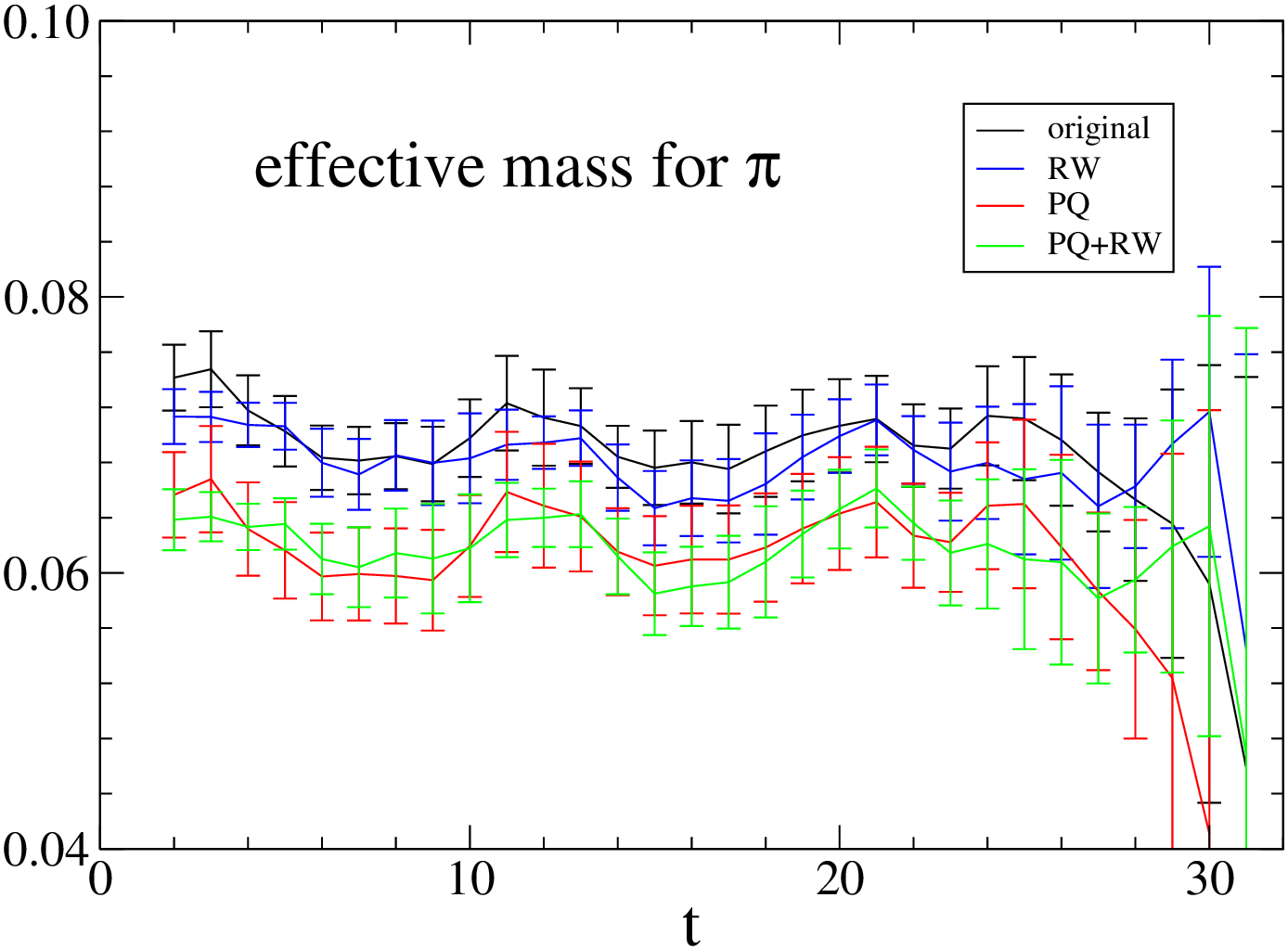}
\hspace*{2mm}&\hspace*{2mm}
\includegraphics[width=85mm,angle=0]{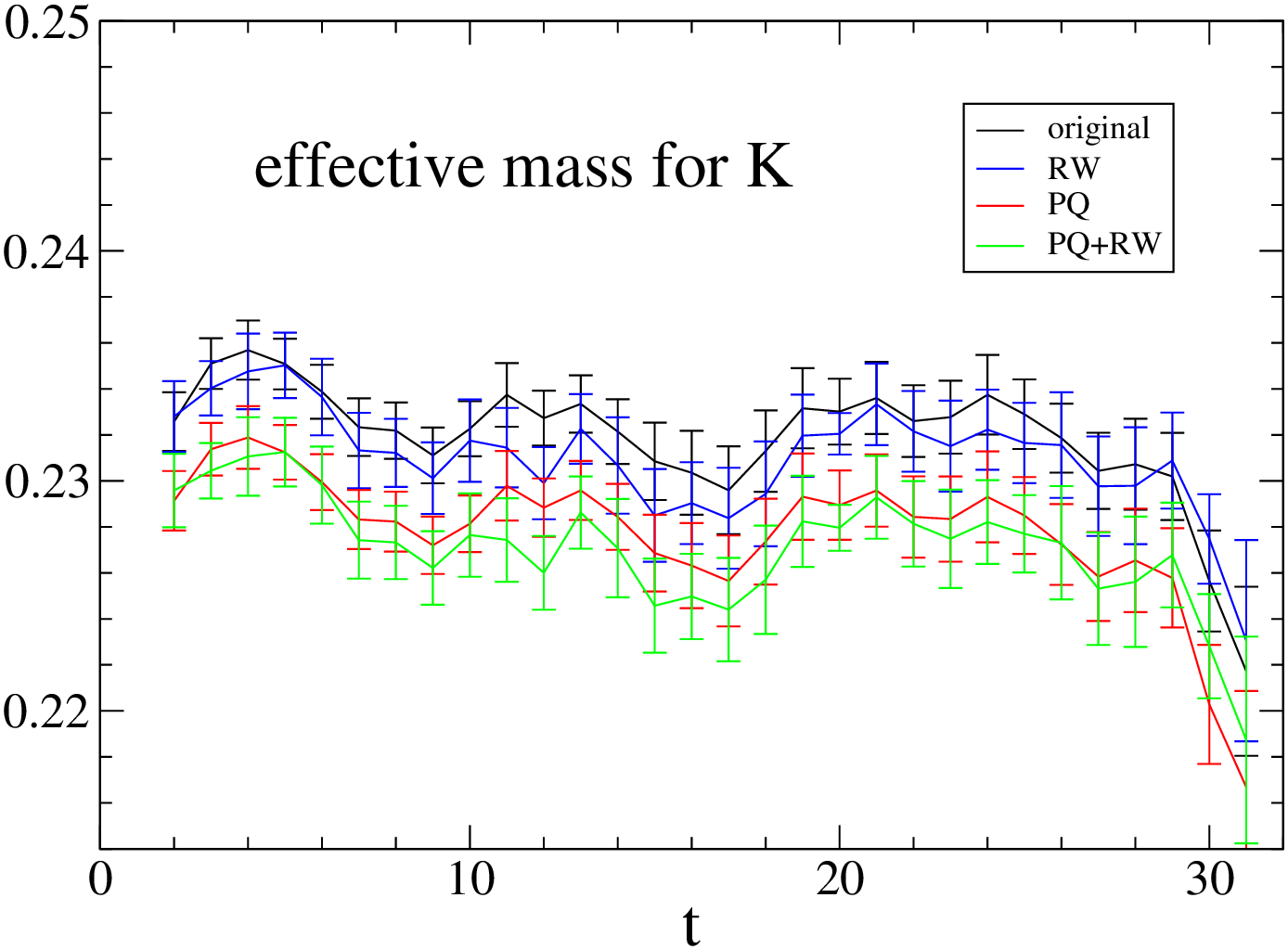}\\
\vspace*{7mm}& \\
\multicolumn{2}{c}
{\includegraphics[width=85mm,angle=0]{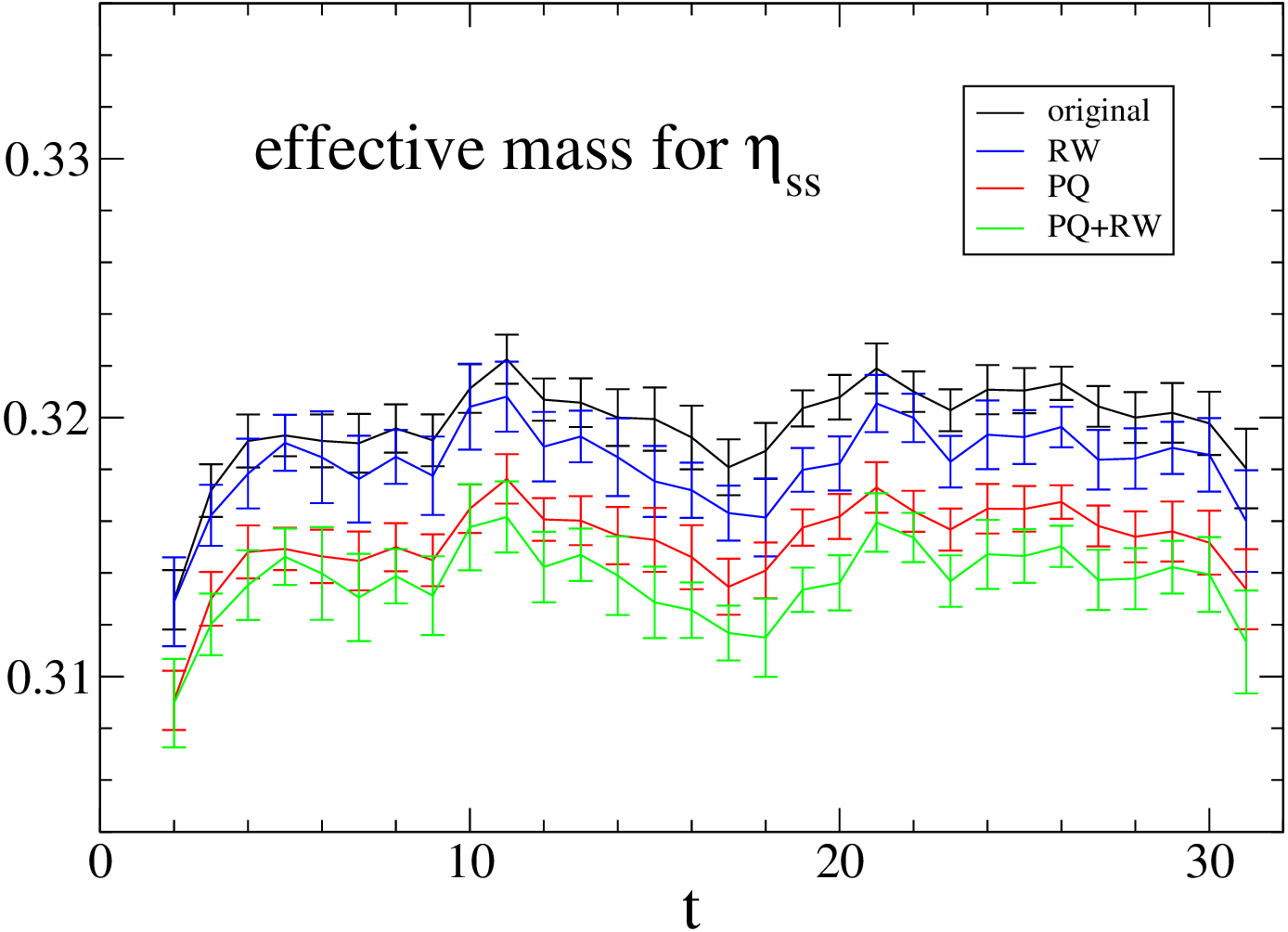}} 
\end{tabular}
\end{center}
\vspace{-.5cm}
\caption{$\pi$, $K$ and $\eta_{\rm ss}$ effective masses with 
the reweighting factors from 
$(\kappa_{\rm ud},\kappa_{\rm s})=(0.137\,785\,00,0.136\,600\,00)$ 
to $(\kappa_{\rm ud}^\target,\kappa_{\rm s}^\target)=(0.137\,796\,25,0.136\,633\,75)$.}
\label{fig:m_eff_rw}
\end{figure*}

To investigate the reweighting effects on the hadron effective masses,
we show the effective masses for the
pseudoscalar mesons with and without the reweighting factors 
in Fig.~\ref{fig:m_eff_rw}, where
$\eta_{\rm ss}$ is a fictitious pseudoscalar meson consisting of
two strange quarks.
For all the cases the partially quenched results (PQ) show lighter 
effective masses than the unitary results at the simulation point.
They are further reduced by the reweighting procedure (PQ+RW).
For other hadron channels the reweighting effects are less clear
partly because of the larger error bars.  
  
\begin{figure*}[h]
\vspace{13mm}
\begin{center}
\includegraphics[width=75mm,angle=0]{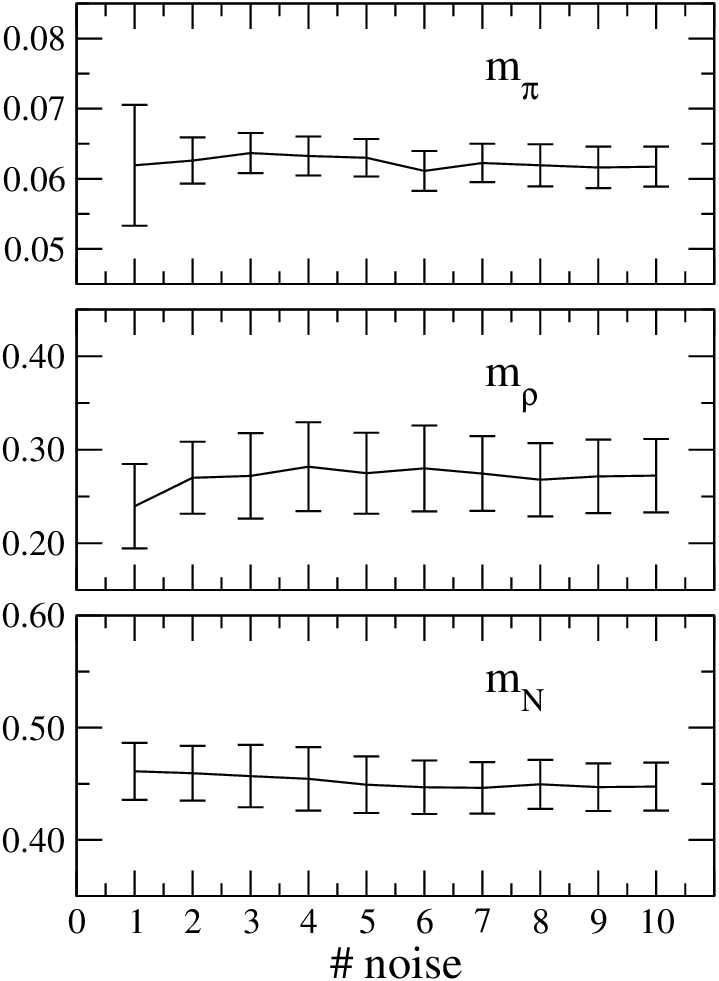}
\end{center}
\vspace{-.5cm}
\caption{$\pi$, $\rho$ and nucleon masses as a function of the number of noise.}
\label{fig:m_rw}
\end{figure*}

In Fig.~\ref{fig:m_rw} we plot the $\pi$, $\rho$ and nucleon masses 
at the target point as a function of the number of noise.
The situation is quite similar to the plaquette case:
Five or six noises appear sufficient to obtain a reliable estimate.
This is also the case for other hadron masses.
   
\begin{figure*}[h]
\vspace{13mm}
\begin{center}
\includegraphics[width=90mm,angle=0]{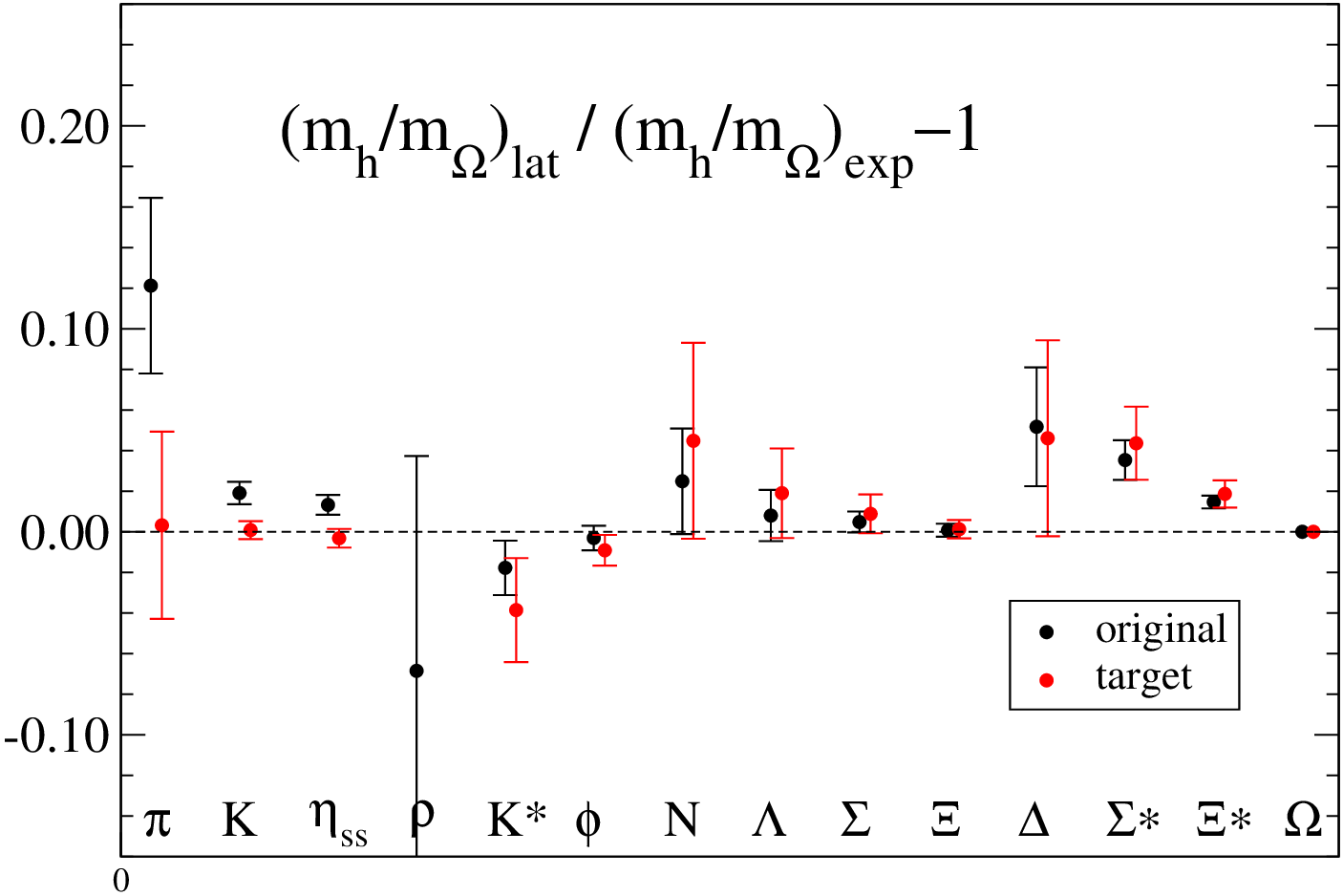}
\end{center}
\vspace{-.5cm}
\caption{Hadron masses normalized by $m_\Omega$ in comparison with
experimental values. Target result for $\rho$ meson locates below the figure.}
\label{fig:m_exp}
\end{figure*}

Figure~\ref{fig:m_exp} compares 
the measured hadron masses normalized by $m_\Omega$ with the 
experimental values.
The results for $m_\pi/m_\Omega$ and $m_K/m_\Omega$, which
are sizably deviated from the experimental values 
at the simulation point (black symbols),
are properly tuned to the physical values within error bars at the
target point of 
$(\kappa_{\rm ud}^\target,\kappa_{\rm s}^\target)=(0.137\, 796\, 25,0.136\, 633\, 75)$.
The lattice spacing  is determined as $a=0.08995(40)$ fm from $m_\Omega$.
A large discrepancy found for $m_\rho/m_\Omega$ may be  
resolved by a proper treatment of $\rho$ meson 
as the resonance\cite{resonance,cppacs_rho}.
We plan to do so for the $\rho$, $K^*$ mesons and $\Delta$ baryon.
For other hadron masses we find less than 5\% deviation from the
experimental values. An increasingly larger deviation observed for lighter
baryons may be due to finite size effects. 

Possible finite size effects on the pseudoscalar meson masses 
based on the NLO
formulae of ChPT\cite{colangelo05} are discussed 
in Sec. IV D of Ref.~\cite{pacscs_nf3}. 
The expected corrections are less than 2\% 
for $m_\pi$ and  $m_K$ at the physical point. 
The magnitude is smaller than the statistical
errors found in Table~\ref{tab:hmass_ph}.
For the baryon masses the heavy baryon ChPT predicts less than 1\%
corrections at the physical point on our physical volume
as listed in Table X of Ref.~\cite{pacscs_hbchpt}.

\begin{figure*}[h]
\vspace{13mm}
\begin{center}
\includegraphics[width=90mm,angle=0]{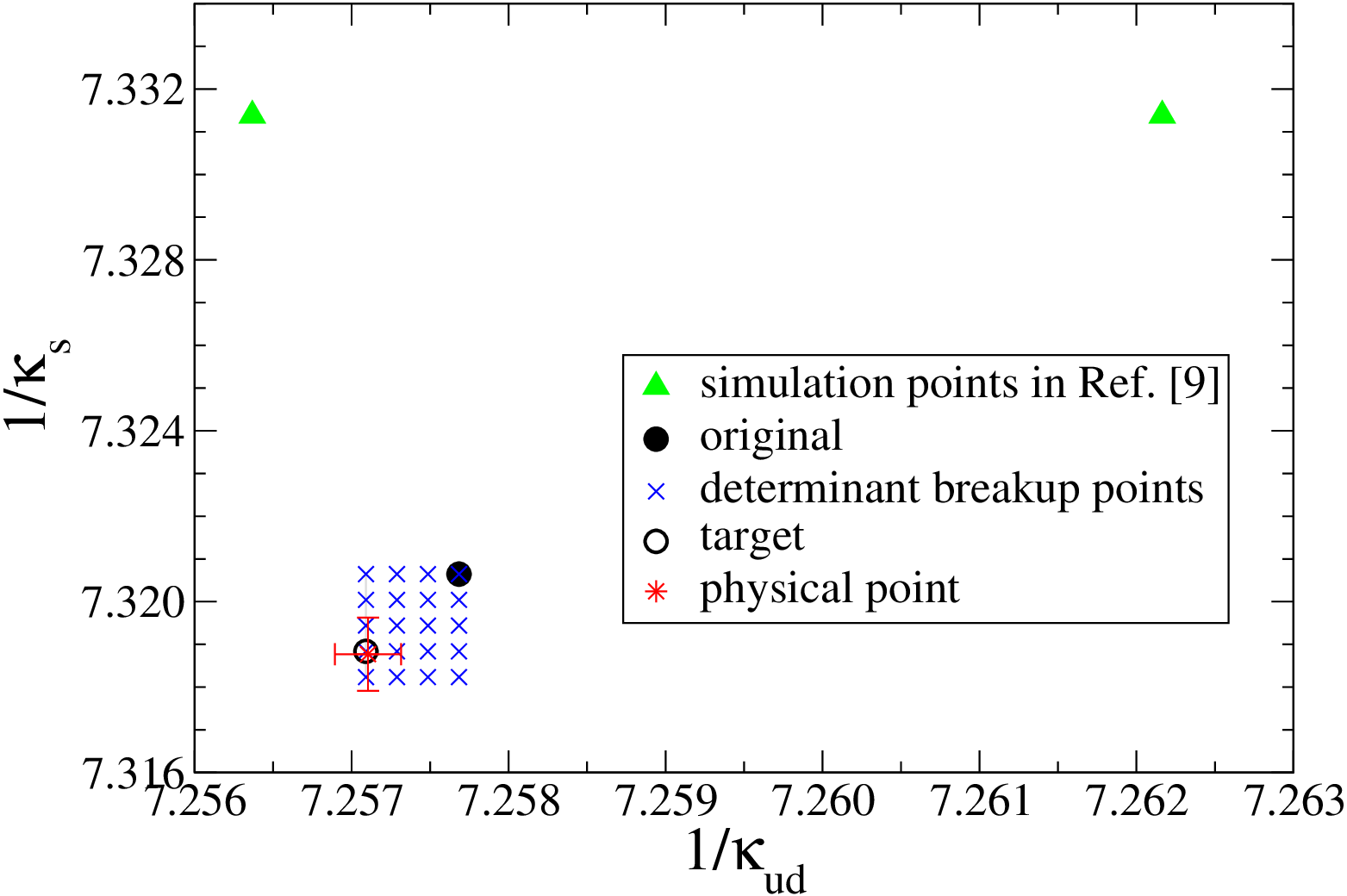}
\end{center}
\vspace{-.5cm}
\caption{Determination of the physical point with $m_\pi/m_\Omega$ 
and $m_K/m_\Omega$ inputs in 
$(1/\kappa_{\rm ud},1/\kappa_{\rm s})$ plane. 
Solid and open black circles denote the original and target points,
respectively.
Green symbols represent
$(\kappa_{\rm ud},\kappa_{\rm s})=(0.137\,810\,00,0.136\,400\,00)$ and 
$(0.137\,700\,00,0.136\,400\,00)$ which are lightest two
simulation points in Ref.~\protect{\cite{pacscs_nf3}}.
}
\label{fig:estimate_pp}
\end{figure*}

Although Fig.~\ref{fig:m_exp} clearly shows that 
further tuning is not really necessary,
it would be instructive to pin down the physical point 
in the $(1/\kappa_{\rm ud},1/\kappa_{\rm s})$ plane.
The physical point plotted in Fig.~\ref{fig:estimate_pp} is
determined by a combined linear fit of $(m_\pi/m_\Omega)^2$ 
and $(m_K/m_\Omega)^2$ at 
$(\kappa_{\rm ud},\kappa_{\rm s})=(0.137\, 785\, 00,0.136\, 600\, 00)$,
$(\kappa_{\rm ud}^{\target},\kappa_{\rm s}^{\target})=(0.137\, 796\, 25,0.136\, 633\, 75)$ 
and two more reweighted points given by
$(0.137\, 796\, 25,0.136\, 633\, 75\pm\Delta_{\rm s})$. 
The fit functions are
\ben
\left(\frac{m_\pi}{m_\Omega}\right)^2&=&c_0^\pi+\frac{c_1^\pi}{\kappa_{\rm ud}}+\frac{c_2^\pi}{\kappa_{\rm s}},\\
\left(\frac{m_K}{m_\Omega}\right)^2&=&c_0^K+\frac{c_1^K}{\kappa_{\rm ud}}+\frac{c_2^K}{\kappa_{\rm s}}
\een
with $c_{0,1,2}^{\pi,K}$ free parameters.
The experimental values of $m_\pi/m_\Omega$ and $m_K/m_\Omega$ are reproduced at $(\kappa_{\rm ud},\kappa_{\rm s})=(0.137\, 797(4),0.136\, 635(16))$, whose central value is almost exactly hit by our target point 
$(\kappa_{\rm ud}^{\target},\kappa_{\rm s}^{\target})=(0.137\, 796\, 25,0.136\, 633\, 75)$. 

The quark masses and the
pseudoscalar decay constants are extracted by repeating the same
analyses as in the simulation
point. The results are summarized in Table~\ref{tab:qmass}.
The quark masses are determined as $m_{\rm ud}^\msbar$(2
GeV)=2.97(28)(03) MeV
and $m_{\rm s}^\msbar$(2 GeV)=92.75(58)(95) MeV
with $a^{-1}=2.194(10)$ GeV, where the second error is due to
the nonperturbative 
renormalization factor obtained by the Schr{\"o}dinger functional 
method\cite{npr_zm}.
We find that our quark masses are comparable to recent estimates in
the literature\cite{qmass_ref}. 
The discrepancy between the quark masses 
in this work and those in Ref.~\cite{pacscs_nf3} is mainly due to the 
difference in the renormalization factors. The nonperturbative estimate
gives about 30\% larger value than the perturbative one\cite{npr_zm}.  
For the pseudoscalar meson decay constants we obtain
$f_\pi=124.1(8.5)(0.8)$ MeV and $f_K=165.5(3.4)(1.0)$ MeV
with the second error coming from 
the nonperturbative renormalization factor\cite{npr_zm}.
These values should be compared with experiment:
$f_\pi=130.4\pm 0.04\pm 0.2$ MeV and $f_K=155.5\pm 0.2\pm 0.8\pm 0.2$ 
MeV\cite{pdg}.
Note that the NLO ChPT analyses predict 4\% (1.5\%) deficit 
for $f_\pi$ ($f_K$) 
on a (3 fm)$^3$ box at the physical point 
due to the finite size effects\cite{colangelo05,pacscs_nf3}.

\section{Conclusion}
\label{sec:conclusion}

We have presented the results of the physical point simulation in 2+1
flavor lattice QCD with the $O(a)$-improved Wilson quark action.
This is accomplished by two algorithmic ingredients: the DDHMC algorithm
with several improvements and the reweighting technique.
The former contributes to cost reduction and the latter is
required for fine-tuning to the physical point. 
 
Clear reweighting effects are observed on several obserbables:
The plaquette value increases and
the hadron masses are reduced in agreement with the expectation for the
reweighting
from the simulation point at 
$(\kappa_{\rm ud},\kappa_{\rm s})=(0.137\, 785\, 00,0.136\, 600\, 00)$ to the target point at
$(\kappa_{\rm ud}^\target,\kappa_{\rm s}^\target)=(0.137\, 796\, 25,0.136\, 633\, 75)$.
We are allowed to properly tune 
the measured values of $m_\pi$, $m_K$ and $m_\Omega$ 
to their experimental ones.

We extract the hadron masses,
the quark masses and the pseudoscalar decay constants directly
on the physical point after the reweighting procedure. 
For the hadron masses we find less than 5\% deviation from the experimental
values except the $\rho$ meson case which requires a proper analysis
as the resonance. The results for the quark masses renormalized in the
$\msbar$ scheme at $\mu=2$ GeV are presented
with the nonperturbative renormalization factor
determined by the Schr{\"o}dinger functional method.
The large enhancement of the quark masses compared to those 
in Ref.~\cite{pacscs_nf3}
is attributed to the difference between the nonperturbative 
renormalization factor and the perturbative one. 

The physical point simulation, which has been 
the long-standing problem in lattice QCD, is achieved in this work.
It appears to us that it is not worthwhile to increase the statistics
with the present simulation parameters.
More important as the next step is 
to repeat the physical point simulation with larger and finer lattices.
Further reduction of the finite size effects and the finite cutoff effects
will make possible precision measurements of physical observables at 1\% level.

\begin{acknowledgments}
Numerical calculations for the present work have been carried out
on the PACS-CS computer 
under the ``Interdisciplinary Computational Science Program'' of 
Center for Computational Sciences, University of Tsukuba. 
A part of the code development has been carried out on Hitachi SR11000 
at Information Media Center of Hiroshima University and 
the INSAM (Institute for Nonlinear Sciences and Applied Mathematics) 
PC cluster at Hiroshima University.
This work is supported in part by Grants-in-Aid for Scientific Research
from the Ministry of Education, Culture, Sports, Science and Technology
(Nos.
16740147,   
17340066,   
18104005,   
18540250,   
18740130,   
19740134,   
20105002,   
20340047,   
20540248,   
20740123,   
20740139    
).
\end{acknowledgments}


\bibliography{apssamp}

\begin{thebibliography}{99}
\bibitem{ukawa1}S. Aoki {\it et al.} (PACS-CS Collaboration), 
Proc. Sci. {\bf LAT2005} (2006) 111.

\bibitem{ukawa2}A. Ukawa {\it et al.} (PACS-CS Collaboration), 
Proc. Sci. {\bf LAT2006} (2006) 039.

\bibitem{boku} T. Boku {\it et al.}, Proceedings of CCGRID 2006 (2006) p. 233. 
\bibitem{csw}
S. Aoki {\it et al.} (CP-PACS and JLQCD Collaborations), 
Phys. Rev. D {\bf 73}, 034501 (2006).

\bibitem{iwasaki}
Y. Iwasaki, Report No. UTHEP-118, 1983 (unpublished).

\bibitem{luscher}
M. L\"uscher, J. High Energy Phys. 05 (2003) 052; 
Comput. Phys. Commun. {\bf 165}, 199 (2005).

\bibitem{massprec1}
M.~Hasenbusch, Phys. Lett. B {\bf 519}, 177 (2001).

\bibitem{massprec2}
M.~Hasenbusch and K.~Jansen, Nucl. Phys. {\bf B659}, 299 (2003).

\bibitem{pacscs_nf3}
S.~Aoki {\it et al.} (PACS-CS Collaboration),
Phys. Rev. D {\bf 79}, 034503 (2009). 

\bibitem{pacscs_hbchpt}
K.-I.~Ishikawa {\it et al.} (PACS-CS Collaboration),
Phys. Rev. D {\bf 80}, 054502 (2009). 

\bibitem{reweighting}
A.M. Ferrenberg and R.H. Swendsen,
Phys. Rev. Lett. {\bf 61}, 2635 (1988). 

\bibitem{cppacs/jlqcd_nf3} 
T. Ishikawa {\it et al.} (CP-PACS and JLQCD Collaborations), 
Phys. Rev. D {\bf 78}, 011502 (2008).

\bibitem{sexton}
J.~C.~Sexton and D.~H.~Weingarten, Nucl. Phys. {\bf B380}, 665 (1992).

\bibitem{kennedy}
A.~Kennedy, Nucl.~Phys.~{\bf B}, Proc. Suppl. {\bf 140}, 190 (2005).

\bibitem{chronological}
R.~Brower, T.~Ivanenko, A.~Levi and K.~Orginos, 
Nucl. Phys. {\bf B484}, 353 (1997).

\bibitem{deflation} 
M.~Parks {\it et al.}, 
SIAM J. Sci. Comput. {\bf 28}, 1651 (2006).   

\bibitem{fastMC}
Ph.~de~Forcrand and T.~Takaishi,
Nucl. Phys. {\bf B}, Proc. Suppl. {\bf 53}, 968 (1997).

\bibitem{Frezzotti:1997ym}
  R.~Frezzotti and K.~Jansen,
    Phys.\ Lett. B {\bf 402}, 328 (1997);
      Nucl.\ Phys.\  {\bf B555}, 395 (1999);
      Nucl.\ Phys.\  {\bf B555}, 432 (1999).

\bibitem{phmc}
S. Aoki {\it et al.} (JLQCD Collaboration), 
Phys. Rev. D {\bf 65}, 094507 (2002).


\bibitem{ishikawa_lat06}
K.-I. Ishikawa {\it et al.} (PACS-CS Collaboration), 
Proc. Sci. {\bf LAT2006} (2006) 27.

\bibitem{Alexandrou:1999ii}  
  C.~Alexandrou, P.~de Forcrand, M.~D'Elia and H.~Panagopoulos,
    Phys.\ Rev. D {\bf 61}, 074503 (2000);
    Nucl.\ Phys. {\bf B}, Proc.\ Suppl.\  {\bf 83}, 765 (2000);
  P.~de Forcrand,
    Nucl.\ Phys. {\bf B}, Proc.\ Suppl.\  {\bf 73}, 822 (1999).

\bibitem{NoisyMetropolisMB}
  A.~Bori\c{c}i and P.~de Forcrand,
    Nucl.\ Phys.\ {\bf B454}, 645 (1995);
  A.~Borrelli, P.~de Forcrand and A.~Galli,
    Nucl.\ Phys.\ {\bf B477}, 809 (1996);
  P.~de Forcrand and A.~Galli,
    arXiv:hep-lat/9603011;
  A.~Galli and P.~de Forcrand,
    Nucl.\ Phys. {\bf B}, Proc.\ Suppl.\  {\bf 53}, 956 (1997).

\bibitem{hasenfratz}
A.~Hasenfratz, R.~Hoffmann and S.~Schaefer,
Phys. Rev. D {\bf 78}, 014515 (2008). 

\bibitem{jung}
C.~Jung, Proc. Sci. {\bf LAT2009} (2009) 002. 

\bibitem{npr_alpha}
S.~Aoki {\it et al.} (PACS-CS Collaboration), 
J. High Energy Phys. 10 (2009) 053.

\bibitem{tani_lat09}
Y.~Taniguchi {\it et al.} (PACS-CS Collaboration), 
Proc. Sci. {\bf LAT2009} (2009) 208.

\bibitem{npr_zm}
S.~Aoki {\it et al.} (PACS-CS Collaboration), 
in preparation.

\bibitem{ca}
T. Kaneko {\it et al.} (CP-PACS/JLQCD and ALPHA Collaborations), J. High Energy Phys. 04 (2007) 092.

\bibitem{pdg}
C.~Amsler {\it et al.} (Particle Data Group), Phys.~Lett.~B {\bf 667}, 1 (2008).

\bibitem{resonance}
M.~L{\"u}scher, Commun. Math. Phys. {\bf 105}, 153 (1986);
Nucl. Phys. {\bf B354}, 531 (1991); {\bf B364}, 237 (1991).

\bibitem{cppacs_rho}
S.~Aoki {\it et al.} (CP-PACS Collaboration), 
Phys. Rev. D {\bf 76}, 094506 (2007).

\bibitem{colangelo05}
G.~Colangelo, S.~D{\"u}rr and C.~Haefeli,
Nucl. Phys. {\bf B721}, 136 (2005).

\bibitem{qmass_ref}
For a recent review, see, E.~Scholz, 
Proc. Sci. {\bf LAT2009} (2009) 005.

\end{thebibliography}

\clearpage

\begin{table*}[h] 
\setlength{\tabcolsep}{10pt}
\renewcommand{\arraystretch}{1.2}
\centering
\caption{Simulation parameters.  
MD time is the number of
trajectories multiplied by the trajectory length $\tau$.
}
\label{tab:param}
\begin{ruledtabular}
\begin{tabular}{ll} 
$\kappa_{\rm ud}$ & 0.137\,785 \\ 
$\kappa_{\rm s}$  & 0.136\,600  \\  \hline 
\#run & 5 \\
$\tau$   & 0.25 \\
block size & $8^4$ \\
$(N_0,N_1,N_2,N_3,N_4)$ &  (4,4,2,4,4) \\
$\rho_1$    & 0.9995 \\
$\rho_2$    & 0.9900 \\
$N_{\rm poly}$ & 220 \\
Replay      & off \\
MD time & 2000 \\
$\la P\ra$ & 0.571\,082(9) \\
$\la {\rm e}^{-dH}\ra$  &  0.9916(81) \\
$P_{\rm acc}$(HMC) & 0.8109(45) \\
$P_{\rm acc}$(GMP) & 0.9519(27) \\
\end{tabular} 
\end{ruledtabular}
\end{table*}

\begin{table*}[h] 
\setlength{\tabcolsep}{10pt}
\renewcommand{\arraystretch}{1.2}
\centering
\caption{Meson and baryon masses in lattice units 
at original and target points.
}
\label{tab:hmass_lat}
\begin{ruledtabular}
\begin{tabular}{llll}
& original & target & physical point in Ref.~\protect{\cite{pacscs_nf3}} \\ 
$\kappa_{\rm ud}$ & 0.137\,785\,00 & 0.137\,796\,25 &  $\cdots$\\ 
$\kappa_{\rm s}$  & 0.136\,600\,00 & 0.136\,633\,75 &  $\cdots$\\  \hline 
$\pi$           & 0.0693(27)  & 0.0617(28)  & 0.0620(9)  \\
$K$             & 0.2321(10)  & 0.2270(9)   & 0.2287(33)  \\
$\eta_{\rm ss}$  & 0.3203(7)   & 0.3138(6)   & 0.3168(43)  \\
$\rho$          & 0.331(38)   & 0.272(39)   & 0.357(16)  \\ 
$K^*$           & 0.4028(55)  & 0.393(11)   & 0.4118(72)     \\
$\phi$          & 0.4652(17)  & 0.4605(28)  & 0.4634(61)     \\ 
$N$             & 0.441(12)   & 0.447(21)   & 0.438(20)     \\ 
$\Lambda$       & 0.5147(63)  & 0.518(10)   & 0.502(10)     \\ 
$\Sigma$        & 0.5485(38)  & 0.5484(62)  & 0.531(11)     \\ 
$\Xi$           & 0.6022(27)  & 0.6001(28)  & 0.5991(75)     \\ 
$\Delta$        & 0.593(16)   & 0.587(27)   & 0.587(19)     \\ 
$\Sigma^*$      & 0.6557(67)  & 0.658(12)   & 0.657(15)     \\ 
$\Xi^*$         & 0.7114(39)  & 0.7113(53)  & 0.718(12)     \\ 
$\Omega$        & 0.7655(34)  & 0.7624(34)  & 0.769(11)     \\ 
\end{tabular}
\end{ruledtabular}
\end{table*} 

\begin{table*}[h] 
\setlength{\tabcolsep}{10pt}
\renewcommand{\arraystretch}{1.2}
\centering
\caption{Meson and baryon masses in physical units 
at target point.
Experimental value for $m_{\eta_{\rm ss}}$ is estimated by
$m_{\eta_{\rm ss}}=\sqrt{2 m_K^2-m_\pi^2}$. 
}
\label{tab:hmass_ph}
\begin{ruledtabular}
\begin{tabular}{llll}
& target [GeV] & physical point in Ref.~\protect{\cite{pacscs_nf3}} [GeV]  & experiment [GeV]\protect{\cite{pdg}} \\ 
$\kappa_{\rm ud}$ & 0.137\,796\,25 &  $\cdots$ & $\cdots$ \\ 
$\kappa_{\rm s}$  & 0.136\,633\,75 &  $\cdots$ & $\cdots$ \\  \hline 
$\pi$           & 0.1354(62)  & $\cdots$  &  0.1350   \\
$K$             & 0.4980(22)  & $\cdots$  &  0.4976   \\
$\eta_{\rm ss}$  & 0.6884(32) & 0.6895(20) &  0.6906   \\
$\rho$          & 0.597(86)  & 0.776(34)  &  0.7755   \\ 
$K^*$           & 0.861(23)  & 0.896(9)   &  0.8960   \\
$\phi$          & 1.0102(77) & 1.0084(40) &  1.0195   \\ 
$N$             & 0.982(45)  & 0.953(41)  &  0.9396   \\ 
$\Lambda$       & 1.137(25)  & 1.092(20)  &  1.1157   \\ 
$\Sigma$        & 1.203(11)  & 1.156(17)  &  1.1926   \\ 
$\Xi$           & 1.3165(60) & 1.304(10)  &  1.3148   \\ 
$\Delta$        & 1.289(59)  & 1.275(39)  &  1.232    \\ 
$\Sigma^*$      & 1.444(25)  & 1.430(23)  &  1.3837   \\ 
$\Xi^*$         & 1.560(10)  & 1.562(9)   &  1.5318   \\ 
$\Omega$        &  $\cdots$       & $\cdots$        &  1.6725   \\ 
\end{tabular}
\end{ruledtabular}
\end{table*}

\begin{table*}[h]
\setlength{\tabcolsep}{10pt}
\renewcommand{\arraystretch}{1.2}
\centering
\caption{Quark masses and pseudoscalar decay constants
at original and target points.
Renormalization factors are nonperturbative in this work,
while perturbative in Ref.~\protect{\cite{pacscs_nf3}}. }
\label{tab:qmass}
\begin{ruledtabular}
\begin{tabular}{lllll}
& original & target & physical point in Ref.~\protect{\cite{pacscs_nf3}} & experiment\protect{\cite{pdg}} \\
$\kappa_{\rm ud}$ & 0.137\,785\,00 & 0.137\,796\,25 &  $\cdots$ & $\cdots$ \\
$\kappa_{\rm s}$  & 0.136\,600\,00 & 0.136\,633\,75 &  $\cdots$ & $\cdots$ \\  \hline
$a{\bar m}_{\rm ud}^{\rm AWI}$    & 0.001\,241(95)  & 0.000\,939(87) & 0.001\,042(32) & $\cdots$ \\
$a{\bar m}_{\rm s}^{\rm AWI}$     & 0.030\,44(9)  &  0.029\,34(12)  & 0.029\,99(70) & $\cdots$ \\
$m^{\overline{\rm MS}}_{\rm ud}$ [MeV]  & 3.92(30)(04) & 2.97(28)(03) & 2.527(47) & $\cdots$ \\
$m^{\overline{\rm MS}}_{\rm s}$  [MeV]  & 96.23(52)(98) & 92.75(58)(95) & 72.72(78) & $\cdots$ \\
$m_{\rm s}/m_{\rm ud}$ & 24.5(1.8)  & 31.2(2.7)  & 28.78(40) & $\cdots$ \\
$af_\pi^{\rm bare}$          & 0.0701(35)  & 0.0661(45)  & 0.0753(22) & $\cdots$ \\
$af_K^{\rm bare}$            & 0.0898(16)  & 0.0881(19)  & 0.0897(18) & $\cdots$ \\
$f_\pi$ [MeV]       & 131.7(6.6)(0.8)  & 124.1(8.5)(0.8) & 134.0 (4.2) & $130.4\pm 0.04\pm 0.2$\\
$f_K$   [MeV]       & 168.7(2.7)(1.0)  & 165.5(3.4)(1.0) & 159.4(3.1) & $155.5\pm 0.2\pm 0.8\pm 0.2$ \\
$f_K/f_\pi$         & 1.280(60)  & 1.333(72)  & 1.189(20) & $\cdots$ \\
\end{tabular}
\end{ruledtabular}
\end{table*}

\clearpage

\end{document}